\documentclass[10pt]{article}
\usepackage{amsmath,psfrag,epsfig,amsfonts}
\usepackage{subfigure}

\DeclareMathOperator{\sutwo}{SU(2)}          
\DeclareMathOperator{\Ad}{Ad}   
\DeclareMathOperator{\tr}{tr}   

\def\be{\begin{equation}}
\def\ee{\end{equation}}
\def\ba{\begin{eqnarray}}
\def\ea{\end{eqnarray}}

\newcommand\nn{\nonumber}
\newcommand\q{\quad}

\newcommand{\eps}{\epsilon}

\makeatletter \@addtoreset{equation}{section} \makeatother

\newcommand{\bd}{\mathbf d}

\newcommand{\Rl}{\mathbb{R}}

\title{Dirac's discrete hypersurface deformation algebras}
\author{Valentin Bonzom, Bianca Dittrich\\
\small Perimeter Institute for Theoretical Physics, Waterloo, Canada}

\date{}

\begin{document}

\maketitle

\begin{abstract}
The diffeomorphism symmetry of general relativity leads in the  canonical formulation to constraints, which encode the dynamics of the theory.  These constraints satisfy a complicated algebra, known as Dirac's hypersurface deformation algebra. This algebra has been a long standing challenge for quantization. One reason is that discretizations, on which many quantum gravity approaches rely, generically break diffeomorphism symmetry. 

In this work we find a representation for the Dirac constraint algebra of hypersurface deformations in a formulation of discrete 3D gravity and for the flat as well as homogeneously curved sector of discrete 4D gravity. In these cases diffeomorphism symmetry can be preserved. Furthermore we present different versions of the hypersurface deformation algebra for the boundary of a simplex in arbitrary dimensions.
\end{abstract}

\section{Introduction}

Diffeomorphism symmetry is the fundamental gauge symmetry of general relativity, deeply entangled with its dynamics. In the canonical framework this gauge symmetry leads to the Hamiltonian and diffeomorphism constraints. They generate multi--fingered time evolution -- that is the equal time hypersurface can be deformed and pushed forward in various ways. These deformations of the hypersurface satisfy a geometric algebra which is reflected in the Poisson algebra of the constraints \cite{teitelboim}, known as Dirac's hypersurface deformation algebra \cite{dirac}.

An important goal for the canonical quantization program is to find a quantum representation of the algebra \cite{habitat,nicolai,vara}. It has not been reached yet, even in the case of 3D gravity which is a topological theory and is well understood in many of its aspects. The Dirac algebra is universal \cite{teitelboim}, i.e. it is the same algebra for any theory of hypersurfaces embedded in a higher dimensional manifold. To understand possible quantum representations of this algebra would therefore be beneficial not only for gravity.

One reason that such a representation is still missing, is that many approaches use discretizations either from the outset or as auxiliary structures to e.g. regularize the quantum constraints. However discretizations generically break diffeomorphism symmetry \cite{loll,dittrichreview,bahrdittrich1}, even classically and in simple systems \cite{steinhaus1,dittrichzako}. For instance, the canonical analysis of Regge gravity \cite{hoehn} shows that in this case constraints are replaced by proper evolution equations \cite{consistent}. In 3D, Regge gravity \cite{regge} does however provide a symmetry preserving discretization, which has to be adjusted to the presence of the cosmological constant \cite{bahrdittrich2} though. Thus we might hope that a quantum representation of the constraint algebra might be obtainable at least in this case. Indeed quantizations of the constraints are available \cite{witten3d,qsd3D,barrettcraneH,valentin1}, although a representation of the Dirac algebra, even classically, has not been found so far.

3D quantum gravity \cite{giddings,thooft,witten,carlip} had been used as a simplified model for 4D gravity for some time. It can be formulated in a number of ways, in traditional (canonical) ADM form, for instance \cite{carlip}, in Chern--Simons form \cite{witten, meusburger} or as a so--called BF theory, which coincides with the Palatini formulation of 3D gravity.

We will work with the latter version, as it is nearest to the variables and methods used in loop quantum gravity, see for instance \cite{notes}. There has been also a considerable amount of work on the loop quantum gravity version of 3D gravity \cite{waelbroeck,qsd3D, alexkarim,PR1,valentin1}. Most of this work, with the exception of \cite{qsd3D} makes use of the fact that the Hamiltonian and diffeomorphism constraints can be recast into the flatness constraints. Whereas the first set satisfies  the Dirac algebra of hypersurface deformations,  the second set is Abelian. 

Despite a good understanding of discrete 3D gravity there is so far no complete discussion or definition of Dirac's hypersurface deformation algebra for discrete surfaces available. The works \cite{bander, utrecht} are incomplete for various reasons, with \cite{bander} arguing (incorrectly) that such an algebra has to be necessarily non--local.

Thus we will provide here an explicit classical realization of the hypersurface deformation algebra for discrete 3D gravity. We discuss the geometric interpretation of the constraints and use these insights to define (various) hypersurface deformation algebras also for discrete 4D gravity (and higher dimensional discretized surfaces). In this case we have however to restrict to the flat sector (or homogeneously curved sector if a cosmological constant is present) of these theories, as otherwise diffeomorphism symmetry will be broken by the discretization. Nevertheless this justifies the hope that some results of the 3D theory can be applied to the 4D case. This might even hold for the quantum theory \cite{valentin4D}.

The organization of the article is as follows. In the Section \ref{sec:continuum}, we review the continuum formulation of 3D gravity, in particular the Dirac algebra, and we present in the Section \ref{3} the discretization we are going to use, inherited from loop quantum gravity. In the Section \ref{4} we introduce the Hamitlonian and diffeomorphism constraints, using a natural space--time splits coming from the loop quantum gravity fluxes. We discuss the form of the algebra and the main tool for its computation in the Section \ref{5}. In the Sections \ref{6} and \ref{7}, we give the explicit Poisson brackets between the constraints associated to a single 3-valent vertex and to a face respectively. The hypersurface deformation algebra is written in terms of lengths and dihedral angles in the Section \ref{8}. The Section \ref{9} shows that the Thiemann trick, used to define the quantum Hamiltonian in loop quantum gravity (in particular \cite{qsd3D}) is also available in the discrete setting, and can be used to write our Hamiltonian constraints. We then move on to higher dimensions, starting with the 4D case in the Section \ref{sec:4Dflat}, in which we argue that a closed algebra is available in the flat sector of the theory. We introduce the corresponding constraints in the Section \ref{simplex}, which work in arbitrary dimension on the boundary of a $d$-simplex. The constraint algebra is found in the Section \ref{sec:bsda}, and generalized to the case of a non-vanishing cosmological constant (describing homogeneously curved space) in the Section \ref{sec:curved}. We close the paper with a discussion in the Section \ref{sec:discussion}. The Appendix \ref{app:calculations} provides details on the evaluation of the Poisson brackets for 3D gravity, relevant for the Sections \ref{6} and \ref{7}, and the Appendix \ref{A} contains background material on the affine metric in a simplex that we use in the Sections \ref{simplex}, \ref{sec:bsda}.

\section{Continuum formulation of 3D gravity} \label{sec:continuum}

3D (Euclidean) gravity can be formulated in first order form, in which the co--triad $e_\mu^k,\, k=1,2,3$ and a connection one--form $A_\mu^k$ are the basic variables. Here $\mu,\rho,\ldots$ denote space time indices and $i,k,j$ internal ($su(2)$ algebra) indices, which are raised and lowered with the internal Euclidean metric $\delta_{kl}$. With the curvature tensor
\ba
F^l_{\mu\nu} &=& \partial_\mu A_\nu^l-\partial_\nu A^l_\mu+\epsilon^{ljk}A_{\mu j}A_{\nu k}
\ea
the action for a 3D manifold ${\cal M}=\Sigma\times \Rl$ can be written as
\be\label{3.2}
S = \frac{1}{2}\int_{\Sigma \times \Rl}   e_{\sigma l}F^l_{\mu\nu} \, \epsilon^{\sigma\mu\nu} \bd^3 x = \int_{\Sigma \times \Rl}  \left( e_b^j \partial_0 A_{aj} \epsilon^{ab} + e_{0j}\tfrac{1}{2}F^j_{ab}  \epsilon^{ab} +  A^j_0(\partial_a e_{bj} + \epsilon_{jlm} A^l_a e_b^m)\tilde\epsilon^{ab} \right) \bd^2 x \bd x^0 .
\ee
Here $\epsilon^{ab},\epsilon^{\sigma \mu \nu}$ are totally antisymmetric Levi--Civita tensor densities and $\epsilon^{ljk}$ is the Levi--Civita tensor in the internal indices. We use the Einstein summation convention. In the second line in (\ref{3.2}) we isolated the time components $\mu=0$ from the spatial components $\mu=a,b$, describing the spatial hypersurface $\Sigma$ by $x^0=\text{const.}$.

Thus we find the momenta
\ba\label{3.3}
E^a_j(x,t)&:=&\frac{\delta S}{\delta (\partial_0A_{a}^j(x,t))}\,=\, \tilde\epsilon^{ab} e_{bj} (x,t)
\ea
 giving the canonically conjugated pair
\ba\label{3.4}
\{A_a^j(x), E^b_k(y)\} &=& \delta^j_k \delta_a^b \delta(x,y) \q .
\ea
The other Poisson brackets vanish: $\{A_a^k(x), A_b^j(y)\}=0$ and $\{E^{a}_k(x),E^b_j(y)\}=0$.

The time components of $e,A$ are Lagrangian multipliers for the action (\ref{3.2}), varying this action with respect to  $e_0^j$ and $A_0^j$ we find the Gau\ss~ and flatness constraints
\ba\label{3.5}
{\cal G}_j&:=& \tfrac{1}{2}T_{abj}  \epsilon^{ab} \,=\, \partial_a E^a_j + \epsilon_{jlm} A^l_a E^{am} \q\q\, \stackrel{!}{=}\;0   \nn\\
{\cal F}^j&:=&\tfrac{1}{2}F^j_{ab} \epsilon^{ab}\; \,=\, \epsilon^{ab}(\partial_aA_b^j +\tfrac{1}{2}{\epsilon^j}_{kl}A_a^kA_b^l)\;\,\;\stackrel{!}{=}\;0 \q
\ea
where $T^l_{\mu\nu}=  \partial_\mu e^l_\nu- \partial_\nu e^l_\mu +\epsilon^{ljk}A_{\mu j}e_{\nu k} - \epsilon^{ljk}A_{\nu j}e_{\mu k}$ is the torsion of the connection $A$ and triad $e$.

These constraints are first class forming the (Galilean symmetry) algebra
\ba\label{3.16}
\{{\cal G}[\Lambda'], {\cal G}[\Lambda]\} &=& {\cal G}[ \,[\Lambda',\Lambda]\,]  \nn\\
\{ {\cal F}[N], {\cal G}[\Lambda]\} &=& {\cal F}[\,[N,\Lambda]\,]  \nn\\
\{ {\cal F}[N'], {\cal F}[N]\} &=& 0  \q
\ea
where ${\cal G}[\Lambda]=\int  \Lambda^j {\cal G}_j d^2 x$ and $ {\cal F}[N] =\int N^j {\cal F}_j d^2x$. The bracket $[A,B]^k=\epsilon^{ijk}A^iB^j$ is the Lie algebra bracket of $su(2)$.

The flatness constraints generate translations in the triad variables $E$ whereas the Gau\ss~constraints generate rotations in the internal space. Gauge symmetries corresponding to diffeomorphisms arise as combinations of the translations and rotations, see for instance  \cite{qsd3D,louapre,notes}.

This leads to the following constraint generating spatial diffeomorphisms
\ba\label{h7}
H_a&=& -e^j_a {\cal F}_j -A^j_a{\cal G}_j \;=\;  E^b_j F^j_{ab} -A^j_a{\cal G}_j \q
\ea
and the Hamiltonian constraint
\ba\label{h8}
H&=& -n^j {\cal F}_j \,=\,-\frac{1}{2} \frac{1}{\sqrt{q}}\epsilon^{jkl}E^a_kE^b_lF_{abj}  \q .
\ea
Here $n^j$ is the normal to the hypersurface $x^0=\text{const.}$
\ba\label{h6}
n^k&=&\frac{1}{2}\frac{1}{\sqrt{\det q}} \tilde \epsilon_{ab} \epsilon^{kjl}E^a_j E^b_l \q .
\ea
This form of the constraints is the same as the form of the constraints in 4D expressed in (self--dual) Ashtekar variables \cite{ashtekarbook}.

The term in (\ref{h7}) proportional to the Gau\ss~constraints is often omitted. Again we introduce the smeared constraints
\ba
H[N] =\int_\Sigma N H d^2x \, ,\q\q H[\vec{N}]=\int_\Sigma N^a H_a d^2 x \q .
\ea
The following Poisson bracket algebra holds modulo terms proportional to Gau\ss~constraints  (see for instance \cite{ashtekarbook,notes} for the computation)
\ba\label{diracalgebra}
\{H[\vec{N}],H[\vec{M}]\} &=& -H[ {\cal L}_{\vec{M}}\vec{N}]    \nn\\
\{ H[N]\,,\, H[\vec{M}]\}&=& - H[ {\cal L}_{\vec{M}} N]  \nn\\
\{H[N]\,,\, H[M]\} &=& + H[ \vec{V}]  \q\q \text{with} \q V^a=q^{ab} \left( M\partial_b N - N\partial_b M\right)  \q .
\ea
Here $q^{ab}$ is the inverse of the two--metric $q_{ab}=e^k_a e_{bk}$, and $\cal L$ is the Lie derivative.

This Poisson algebra is known as \emph{Dirac's hypersurface deformation algebra} and describes the commutator of normal and tangential deformations of a hypersurface. It is universal, i.e. holds in all dimensions and with arbitrary field content. A special feature of the Dirac algebra is the appearance of the structure function $q^{ab}$ in the brackets between two Hamiltonian constraints.

\section{Discretization of 3D gravity}\label{3}

Here we will review shortly the standard choice of discretization for 3D gravity in first order form, \cite{waelbroeck, PR1, notes}.

The internal index $k$ in $A_a^k, e_b^k$ transforms according to the fundamental representation of $SO(3)$, i.e. the spin one representation. This agrees with the adjoint representation on the Lie algebra $su(2)$, which suggest to introduce the  Lie algebra valued forms $A_a = A^k_a T^k$ and  $e_b=e_b^k T^k$, where $T^k$ is a basis of $su(2)$.
 We will work with $T_k=T^k=-\frac{i}{2}\sigma_k$ where  $\sigma_k$ are the Pauli matrices.\footnote{These satisfy $\sigma_j\sigma_k=\delta_{jk} \mathbf{1} + i {\epsilon_{jk}}^l\sigma_l$, from which $T_iT_j =-\tfrac{1}{4} \delta_{ij}\mathbf{1}+ \tfrac{1}{2} {\epsilon_{ij}}^kT_k $ and $T_iT_jT_m  = \tfrac{1}{4} \left(   \delta_{im}T_j  -\delta_{ij} T_m    -  \delta_{jm}T_i     \right)   -        \tfrac{1}{8} \epsilon_{ijm} \mathbf{1} $ follows.}

The parallel transport along a curve $\gamma$ of a Lie algebra valued object $V$ is defined by $V(s)=h(s)V(0)h(s)^{-1}$ where $h(s)$ is the  holonomy of the connection $A$ given by
\ba\label{hol11}
h_\gamma(s)&=& \mathbb{P} \exp\left[-\int_\gamma A\right] \nn\\
&:=&
\sum_{n=0}^\infty (-1)^n \int^s_0  \bd s_n \int^{s_n}_0 \bd s_{n-1} \cdots \int^{s_{2}}_0 \bd s_1 \, A(s_n) A(s_{n-1}) \cdots A(s_2)A(s_1) \q\q
\ea
with $A(s)=\dot\gamma^a(s) A_a(\gamma(s))$. A dot denotes differentiation with respect to the curve parameter $s$. Note that with this definition we have for the composition of edges $e_1 \circ e_2$ the relation $h_{e_1\circ e_2}=h_{e_2} h_{e_1}$.

Our choice of discretization will replace the connection by holonomies $g_e:=h_e(1)$ along oriented edges $e$. The triad variables will be encoded into the fluxes
\ba\label{fl1}
E_{e} &=& \int_{e^*}  h_{e,e^*}(s) \, e_a(e^*(s) )(\dot{e^*})^a(s) \, \left( h_{e,e^*}(s)\right)^{-1} \,\bd s \q .
\ea
where $e^*$ is an edge cutting $e$ transversally and such that $(e,e^*)$ are positively oriented.
Here $h_{e,e^*}(s)$ parallel transports vectors from the point $s$ on $e^*$ to the source vertex $v_s=e(0)$ of $e$. We define $h_{e,e^*}(s)$ as follows:  Choose the parametrization of the curves $e$ and $e^*$ such that the intersection point in both cases corresponds to $s=\frac{1}{2}$.  Then
\ba\label{ppp}
h_{e,e^*}(s) &=&(h_e(\tfrac{1}{2}))^{-1} \cdot  h_{e^*}(\tfrac{1}{2},s)
\ea
where $h_{e^*}(\tfrac{1}{2},s)$ is the parallel transport from the parameter $s$ to the parameter $\tfrac{1}{2}$ along $e^*$.

We inserted the holonomies to ensure a local transformation behaviour of the fluxes under internal rotations
$E_{e}  \rightarrow   g(v_s)  \,E_{\gamma} \, g(v_s)^{-1}$
where $g(\cdot)$ denotes the gauge field on the spatial hypersurface $\Sigma$.

 Under the inversion of the orientation of the edge $e \rightarrow \bar{e}:=e^{-1}$  the holonomies and fluxes transform as
\ba\label{fl3}
h_{\bar{e}} \;=\; (h_{\bar e})^{-1} \q ,\q\q
 E_{\bar{e}} \; =\; -h_e E_{e} (h_e)^{-1} \q .
 \ea

The Poisson brackets between holonomies and fluxes can be computed and result \cite{qsd7,notes} in
\ba\label{fluxhol}
 \{(g_e)_{MN}, (g_{e'})_{M'N'}\} &=& 0 \q ,  \nn\\
\{ E^j_e\, , g_{e'} \}&=& \delta_{ee'}g_eT^j \q   \nn\\
 \{ E^j_e, E^k_{e'} \}&=&\delta_{ee'} {\epsilon^{jk}}_l E^l_\gamma \q .
\ea
We can deduce from (\ref{fluxhol}) for the bracket with an inverse group element
$
\{ E^j_e\, , g^{-1}_{e} \}\,=\, -T^j g_e^{-1}
$
as well as the relation
$
\{E_{\bar{e}}^k, E^l_e\}=0     .
$

The Gau\ss~constraints are assigned to vertices and are represented as
\ba
G_v &=&  \sum_{ e: v_s(e)=v } E_{e} \;+\; \sum_{e: v_t(e)=v} E_{\bar{e}} \q ,
\ea
i.e. as sum over the fluxes associated to the outgoing ($e: v_s(e)=v$) and the incoming $e: v_t(e)=v$ edges at $v$. The Gau\ss~ constraints generate gauge transformations on the internal index, i.e. internal rotations. Later--on, in the Section \ref{8}, we will discuss the reduced phase space with respect to the Gau\ss~ constraints.

From the construction of the holonomy and flux variables we can infer the following interpretation. The flux $E_e$ gives the components of the edge $e^*$ in a frame associated to the source vertex $v_s(e)$ of $e$. The holonomy variables $g_e$ are used to transport between the frames associated to the source vertex $v_s(e)$ and target vertex $v_t(e)$ of $e$. Thus the Gau\ss~constraints impose that the fluxes around a vertex sum to zero, which geometrically imposes that the dual edges $e^*$ form a closed curve. For a three-valent vertex this will define a closed triangle. (For higher--valent vertices the edges $e^*$ making up the piecewise linear curve close, but will in general not span a 2--dimensional subspace.) Due to this interpretation we will  assume that the edges and vertices are the elements of a 2--complex, which discretizes the 2D surface $\Sigma$. This latter fact means that we can also identify faces $f$. The 2--complex is usually assumed to arise as the dual of a triangulation (in which case the vertices are three--valent) or more generally a discretization with polygonal cells (which allows higher valent vertices).

The flatness constraints impose that the parallel transport around (contractible) loops is trivial. The most elementary loops are given by the boundaries of the faces. We have furthermore to specify the frame, i.e. vertex, in which the holonomy is expressed. Thus we define the flatness constraints\footnote{We choose a symmetrized form of the constraints, as this is the form that is usually employed in the Hamiltonian constraints. It however also allows for the solution $h_{fv}=-\mathbb{I}$. One can also use $F_{fv}=\mathbb{I}-h_{fv}$ -- this will not change the algebra of the constraints. In the following we will understand as constraint hypersurface the subspace defined by $h_{fv}=\mathbb{I}$. } as
\be
F_{f,v}:=\tfrac{1}{2}(h^{-1}_{fv}-h_{fv})
\ee
with  $h_{fv}$  the holonomy around a face (dual to a vertex in the triangulation) starting with the edge $e \subset f$ which we assume to have $v$ as a source vertex $v=v_s(e)$, i.e. $h_{fv} = \cdots g_{e''} g_{e'} g_{e}$. In the following, we will often parametrize $\sutwo$ matrices as $h =h^0\mathbb{I} + h^i T^i$, where $h_i =-2\tr (hT^i)$ is the projection of $h$ onto the generator $T^i$. The inverse of $h$ writes $h^{-1} = h^0\mathbb{I}-h^i T^i$. Therefore,
\be
F_{f,v} = - h^i_{fv}\,T^i.
\ee

As can be easily checked the algebra of flatness and Gau\ss~ constraints is again first class, with the flatness constraints forming an Abelian subalgebra.  In the following subsection we will find combinations of these constraints, that can serve as discrete versions of the Hamiltonian and diffeomorphism constraints.

\section{Geometry and constraints associated to a face vertex pair}\label{4}

The action of the flatness constraints on the flux associated to $e$ is given by
\be\label{acf}
\{ E^k_e, F_{f,v}\} \,=\,- \frac{1}{2} \bigl(T^k h^{-1}_{fv}  + h_{fv} T^k\bigr) \simeq -T^k \q
\ee
with the last equation holding on the constraint hypersurface. Again we assume that $h_{f,v} = \cdots g_{e''} g_{e'} g_{e}$. The action (\ref{acf}) of the flatness constraints generalizes to the fluxes $E_{e'}$ with $e'$ in the boundary of $f$, if these are transported to the reference frame of $v_s(e)$. Thus  $F_{f,v}$ generates translations in the (to $v$ parallel transported) components $E^k_{e'}$ of the fluxes associated to the edges $e'$ in the boundary of the face $f$. This face $f$ is dual to a vertex $f^*$ in the dual triangulation, and we assume that the face is such that $e^*$, the dual edge to $e$, points towards the dual vertex $f^*$.

Now contract $F_{f,v}$ with a vector $-n^k$,
\be
-F_{f,v} \cdot n := -\sum_l F^l_{f,v} n^l \,:=\,- \sum_k (-2) \tr (F_{f,v} T^l n^l ) \,\, .
\ee
This defines a combination of constraints which according to
\be
\{ E^k_e, -F_{f,v} \cdot n\} \, \simeq \, \sum_l (-2) \tr(  T^k  T^l n^l) = n^k \q
\ee
generates the translation of $E^k_e$ by the vector $n^k$.

Thus Hamiltonian (and diffeomorphism) constraints may be defined by contracting the holonomy around each face with some choice of normals (and vectors tangent to the hypersurface). However, the holonomy $h_{fv}$ probes the curvature around a vertex of the triangulation and there is no obvious notion of tangent and normal vectors to the vertex. One could proceed with some averaging over the normals of the adjacent triangles. That would require to specify the details of the averaging. The loop quantum gravity version \cite{qsd3D} proceeds in another way and associates one Hamiltonian constraint (the diffeomorphisms being treated with a different method) to each dual vertex (corresponding to a triangle in the triangulation). Thus there is an averaging over the three vertices of the triangle and correspondingly of the three holonomies involved. Note that the counting of constraints might not necessarily match up in this case, that is one might obtain a redundant set or an incomplete set. For 4D gravity this discrepancy has been noted in \cite{immirzi}, here we see that one should be aware of possible redundancies between the constraints.

Instead of averaging, we will work with an over--complete basis of constraints. Possible averagings can be performed afterwards and the constraint algebra adjusted, which in particular is straightforward if the averaging coefficients are constants.

\begin{figure}
\center
\includegraphics[scale=.5]{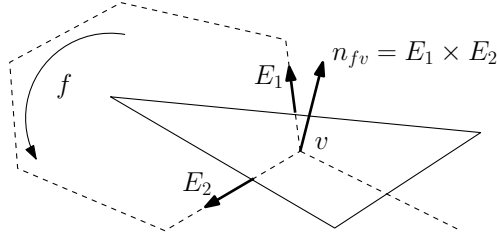}
\caption{\label{fig:LocalFrame} The vertex $v$ is dual to the triangle (in solid lines) and the face $f$ (dashed lines) is dual to a vertex of the triangle. $E_1, E_2$ are associated to the two edges meeting at $v$ in the boudary of $f$. They describe the embedding of the triangle in $\mathbb{R}^3$, with normal $n_{fv}=(E_1\times E_2)/|E_1\times E_2|$.}
\end{figure}

A vertex of the triangulation is shared by at least three triangles. Each of them has a local frame, a basis of which being provided by the two edge vectors of the triangle meeting at the vertex, say $E_1, E_2$, which span the tangential directions to the triangle, and an obvious choice of normal,
\be \label{normal}
n_{fv} = (E_1\times E_2)/|E_1\times E_2|.
\ee
In the dual graph, the edge vectors $E_1, E_2$ are equivalently identified by a pair face-vertex $(f,v)$. The face $f$ is dual to the vertex of the triangulation, and the vertex $v$ labels the triangle that provides a local basis. $E_1,E_2$ are the vectors associated to the two dual edges meeting at $v$ in the boundary of $f$. In the definition of the normal \eqref{normal}, we have assumed that $E_1, E_2$ are both outgoing (or both ingoing) at $v$. The situation is summarized in the Figure \ref{fig:LocalFrame}.

We can now define a Hamiltonian and diffeomorphism constraints by contracting the holonomy, or rather the flatness constraint $F_{f,v}$  with the appropriate vectors for each pair $(f,v)$, 
\be
\begin{aligned}
H^{fv} &=  -F_{fv}\cdot n_{fv} = n^k\, h^k_{fv} = \frac{(E_1\times E_2)^k}{|E_1\times E_2|}\, (-2) \tr(h_{fv} T^k),\\
D^{fv}_\alpha  &= -F_{fv} \cdot E_\alpha = E_\alpha^k\, h^k_{fv} = E_\alpha^k\,(-2)\tr(h_{fv} T^k) \q.
\end{aligned}
\ee
Here and in the following we will sum over pairs of internal indices, even if these are not in the usual summation convention position (as the internal indices are raised and lowered with the trivial metric $\delta_{kl}$).

To describe the algebra generated by those constraints, it will be very convenient to use the following metric per pair $(f,v)$,
\be
Q^{fv}_{\alpha\beta} = E_\alpha \cdot E_\beta \, , \qquad \alpha,\beta=1,2.
\ee
Its determinant is $\det Q = |E_1\times E_2|^2$, so that when $E_1, E_2$ are linearly independent, it has an inverse,
\be\label{qinv}
Q^{-1}_{fv} = \frac{1}{|E_1\times E_2|^2}
\begin{pmatrix}
E_2^2 & - E_1\cdot E_2 \\
-E_1\cdot E_2 & E_1^2
\end{pmatrix}\;,
\ee
whose matrix elements will be denoted $Q_{fv}^{\alpha\beta}$. Note that this really is the analogue of the continuum inverse 2-metric, which we expect to appear in the Dirac algebra. It naturally induces a `space--time split' as
\ba\label{split}
\delta^{ij} &=& E^i_\alpha \, Q_{fv}^{\alpha \beta}\,  E^j_\beta + n_{fv}^i n_{fv}^j  \q .
\ea
Here we sum over $\alpha, \beta=1,2$. This identity can be used to decompose any vector $A$ into components tangential to the triangle dual to $v$ and parallel to the normal $n$. In the case of the flatness constraint/holonomy $h^i_{fv}$, the components are exactly the diffeomorphism and Hamiltonian constraints,
\be
h^i_{fv} = E_\alpha^i\,Q^{\alpha \beta}_{fv}\,D_\beta^{fv} + n^i\,H^{fv}.
\ee

The constraints are redundant in two ways:
\begin{itemize}
\item
Local redundancy: The holonomy has three real degrees of freedom, so that the flatness constraints give three constraints per face. We have define three constraints for every face--vertex pair $(f,v)$, which if the two edge vectors associated to $(f,v)$ are linear independent, are also independent. Thus in general the constraints associated to different vertices at the same face can be related to each other.
\item
Global redundancy: The constraints generate vertex translations of the triangulation (which can be locally embedded into 3D flat space time due to the equations of motion). Given for instance a spherical triangulation, we can translate all vertices at once such that we either produce a global rotation or a global translation. Thus in this case we have six combinations among the vertex translations, which do not change the length and the dihedral angles of the triangulation. The latter provide a parametrization of the Gau\ss~constraint reduced phase space.   Thus for the boundary of a tetrahedron one will have 4 (triangulation) vertices, hence 12 (flatness) constraints. Only six of these are independent, fixing the dihedral angles as functions of the lengths \cite{dittrichreview}.   Related is the redundancy between the flatness constraints (on the full phase space) induced by the integrated Bianchi identity. For a spherical surface this identity implies that the holonomy of a given face can be written as a combination of the holonomies of other faces.
\end{itemize}
The latter global redundancy has been overlooked in \cite{bander} and lead there to the conclusion that it would not be possible to obtain a closed (local) hypersurface deformation algebra for discrete geometries. We will show that this is clearly not the case.

\section{The constraint algebra}\label{5}

The form of the continuum constraint algebra (\ref{diracalgebra}) can actually be derived by geometrical considerations \cite{teitelboim} -- it reflects the commutator of deformations of a hypersurface embedded in a higher dimensional manifold. This shows that the algebra is universal, i.e.\ it is independent from the field content. Moreover, given a certain set of assumptions on the number and metric interpretation of the (ADM) phase space variables, the Hamiltonian and diffeomorphism constraints can be derived uniquely \cite{teitelboim2}. (This work has been performed in the ADM variables, we are not aware of a derivation in connection or Ashtekar--Barbero variables.)

We will also derive the constraint algebra by geometric considerations, for the boundary of a simplex of arbitrary dimensions, in section \ref{simplex}. Let us shortly explain how the form of the algebra arises.
\begin{itemize} \parskip -2pt
\item[(a)]
The commutator of diffeomorphism constraints with diffeomorphism constraints gives again diffeomorphism constraints. This is due to the diffeomorphisms describing deformations in the tangent space to the spatial hypersurface. The commutator does not leave this hypersurface.
\item[(b)]
The commutator of a Hamiltonian and a diffeomorphism constraint gives a Hamiltonian. The normal to the hypersurface does not change under tangential deformations of this hypersurface. However a tangential vector defining the spatial diffeomorphism changes under a normal deformation -- namely by the normal itself. Thus we obtain a Hamiltonian deformation as the commutator.
\item[(c)]
In (\ref{nderiv}) we derived the variation of the normal under the change of hypersurface (in this case described by the fluxes). Due to the normalization of $n$ the variation is orthogonal to the normal itself, that is we obtain a tangential deformation. Moreover we have the inverse spatial metric appearing in (\ref{nderiv}) which explains why it appears as a structure function.
\end{itemize}

This will hold in the discrete context as well, but only for the Poisson brackets between two constraints for which the space--time splittings \eqref{split} are with respect to one and the same normal. This is the case for constraints around a single 3-valent vertex (due to the Gau\ss~constraint), as we will see in the Section \ref{6}. However, when computing the algebra around a face, in the Section \ref{7}, involving different vertices, hence different spaced--time splits, we will see that the brackets are a bit more complicated. Nevertheless, they remain geometrically transparent, the point being that a tangential (or normal) deformation to a triangle dual to a vertex $v$ is typically not tangential (or normal) when seen in the local space--time split of a different triangle, say dual to the vertex $v'$. Therefore, a spatial deformation at $v$ has to be non-trivially decomposed into diffeomorphisms and Hamiltonian at $v'$ (and the same for normal deformations).

Moreover, those geometric considerations only apply to the part of the algebra which is linear in the constraints as it refers to the geometric action that the constraints generate (we will see that the brackets involve terms quadratic in the constraints, which however do not generate an action on the constraint hypersurface).


We consider now the Poisson algebra between two constraints. They can in general be associated to two different faces $f,f'$ and vertices $v,v'$. We will denote the (outgoing) edges associated to $(f,v)$ by $e_1,e_2$ (with indices $\alpha,\beta=1,2$) and the ones associated to $(f',v')$ by $e_3,e_4$ (with primed indices $\alpha',\beta'=3,4$ indices). One or both edges $e_3,e_4$ might coincide or be an inverse of one or both of the edges $e_1,e_2$.  However, only if $v=v'$ we will have that the fluxes associated to these edges might not commute. Also we need to have $v\subset f'$ or $v' \subset f$  (or both) for the Poisson brackets not to vanish, in addition the two faces need to share an edge.

The constraints arise as contractions of the face holonomy with either the fluxes or the normal. Let us denote this choice by $A$ for $(f,v)$ and by $B$ for $(f'v')$. The Poisson brackets can then be written as
\begin{multline}\label{pb1}
\{A^l h^l_{fv} \, ,\, B^m h^m_{f'v'}\} =
h^l_{fv}   \frac{\partial A^l}{\partial E_\alpha^k} \,\{ E_\alpha^k, h^m_{f'v'}\}  \,B^m \;-\;
h^l_{f'v'}  \frac{\partial B^l}{\partial E_\gamma^k} \, \{ E_\gamma^k, h^m_{fv}\} A^m
\\+\, \delta_{vv'} h^l_{fv} \, h^m_{f'v'}   \frac{\partial A^l}{\partial E_\alpha^k}   \frac{\partial B^m}{\partial E_\gamma^p} \{ E_\alpha^k\, ,\, E_\gamma^p\}  \q .
\end{multline}

Notice that all terms on the right hand side of this equation vanish if $h^l_{fv}=h^m_{f'v'}=0$, thus the constraint algebra still closes (as expected). 

For $A= E_{\beta}^i$, the derivative is $\partial E_\beta^i/\partial E_\alpha^k = \delta_{\alpha,\beta}\delta^{ik}$. We also need the derivative of the normal with respect to the edge vectors. This can be computed explicitly, but also found from the equations
\ba
\frac{\partial}{\partial E_\alpha^k} \,( n\cdot n) &=& 2 n^l \frac{\partial n^l}{\partial E_\alpha^k} \;=\;0 \nn\\
\frac{\partial}{\partial E_\alpha^k} \,(n\cdot E_\beta) &=& E_\beta^l \frac{\partial n^l}{\partial E_\alpha^k}   + n^k \delta^\alpha_\beta \;=\; 0 \q .
\ea
Taking the contraction of these equations with $Q^{\beta\gamma}E_\gamma^m$ we conclude
\ba\label{nderiv}
 \frac{\partial n^l}{\partial E_\alpha^k}  &=& -n_k \, Q^{\alpha\beta} E_\beta^l  \q .
\ea

The main ingredient to evaluate all the terms in \eqref{pb1} is the bracket between a flux at $v$ and a holonomy component for $(f',v')$. We introduce the notation $h_{f':vv'}$ for the holonomy from $v$ to $v'$ along the $f'$ (using its orientation) and $h_{f':v'v}$ for the holonomy from $v'$ to $v$. Hence $h_{f'v'}= h_{f':vv'} h_{f':v'v}$ and $h_{f'v}= h_{f':v'v} h_{f':vv'}$. The holonomy $h_{fv'}$ goes along the edge $e_\alpha$ and hence contains $g_\alpha^{o_{f\alpha}}$, where $o_{f\alpha}=\pm$ is the relative orientation. The bracket between $E_\alpha^k$ and $h_{f'v'}$ inserts the generator $T^k$ at $v$, the global sign being $o_{f\alpha}$. Thus we can write
\ba\label{pb2}
\{ E_\alpha^k, h^m_{f'v'}\}
&=& o_{\alpha f'}  \,(-2) \tr ( h_{f':vv'} T^k \,  h_{f':v'v} \, T^m)\,\} \nn\\
&=& o_{\alpha f'}  \,(-2) \tr (   h_{f':v'v}  h_{f':vv'}  \,T^k \,   \text{Ad}_{f':v'v}(T^m) ) \nn\\
&=& o_{\alpha f'}  \,(\Ad_{f':v'v}(T^m) )^n \,(-2) \tr (   h_{f'v} \,T^k   \,T^n)\;.
\ea
Note that $o_{\alpha f'}$ can be set to $0$ if $e_\alpha$ is not in $f'$. Also $ \Ad_{f':v'v}(B)$ denotes the adjoint of $h_{f':v'v}$ on the vector $B$ (which is in the reference frame of $v'$), i.e. $\Ad_{f':v'v}(B)=  h_{f':v'v} \, B \, h^{-1}_{f':v'v}$. This is the transport of $B$ to the reference frame of $v$.

We need to be careful with the case $v=v'$, as then in the last line of \eqref{pb2} either $h_{f':vv'}=h_{f'v}$ and $h_{f':v'v}=\mathbb{I}$ (for $o_{\alpha f'}=1$), or the other way around (for $o_{\alpha f'}=-1$). In the latter case we have
\ba
\{ E_\alpha^k, h^m_{f'v'}\}
&=&   o_{\alpha f'}  \,(-2) \tr (   h_{f'v} \,T^k   \, (\text{Ad}_{f':v'v}(T^m) )^nT^n) \nn\\
&\underset{ v=v'\,\,,o_{\alpha f'}=-1 }{=}& -\,(-2) \tr (   \,T^k  h_{f'v} T^m  )  \q .
\ea
To evaluate the trace in the last line of (\ref{pb2}) we expand $h_{f'v}=h^0_{f'} \mathbb{I} + h^p_{f'v} T^p$. We note that $h^0_{f'} =\frac12 \tr h_{f'v}$ is actually independent of the vertex $v$. Thus,
\ba\label{pb3}
\{ E_\alpha^k, h^m_{f'v'}\}
&=&
o_{\alpha f'} h^0_{f'} (\Ad_{f':v'v}(T^m) )^k  \,+\,  \tfrac{1}{2} o_{\alpha f'} \, h^p_{f'v}   \, \epsilon^{pkn} \,  (\Ad_{f':v'v}(T^m) )^n \q , \nn\\
\{ E_\alpha^k, h^m_{f'v'}\}&\underset{ v=v' }{=}& o_{\alpha f'}   h^0_{f'} \delta^{km} \,+\, o^2_{\alpha f'} \tfrac{1}{2}  \, h^p_{f'v}   \, \epsilon^{pkn}   T^n \q .
\ea
In the last line we use the notation $o^2_{\alpha f'}=1$ if $\alpha \subset f'$ and  $o^2_{\alpha f'}=0$ otherwise. 

The evaluation of all relevant brackets, for the generic case with $f\neq f'$ and $v\neq v'$, is derived in the Appendix \ref{app:calculations}.

\section{Constraints at a vertex}\label{6}

\begin{figure}[hbt]
\begin{center}
\psfrag{1}{$e_1$}
\psfrag{2}{$e_2$}
\psfrag{3}{$e_3$}
\psfrag{4}{$e_4$}
\psfrag{5}{$e_5$}
\psfrag{6}{$e_6$}
\psfrag{v}{$v$}
\psfrag{f}{$f$}
\psfrag{fp}{$f'$}
 \includegraphics[scale=0.5]{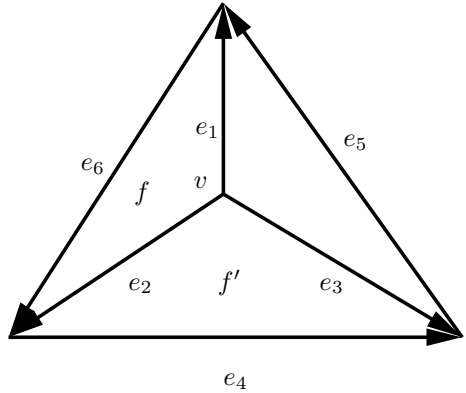}
 \caption{The vertex $v$ is 3--valent, hence surrounded by three faces, and the edges $e_1, e_2, e_3$ meet at $v$.} \label{tetra}
 \end{center}
\end{figure}

We consider a three-valent vertex $v$, dual to a triangle of the triangulation. It is shared by three faces $f, f', f''$, so we have three flatness constraints. Since the triangle is closed, the flux variables meeting at $v$ satisfy the Gau\ss~constraint, $E_1+E_2+E_3=0$, assuming the edges are all outgoing. A priori we have three different normals $n_{fv},n_{f'v},n_{f''v}$ at $v$. However they all coincide modulo terms proportional to the Gau\ss~constraint and describe the (unique) normal of the triangle dual to $v$. The following Poisson brackets are therefore modulo terms proportional to the Gau\ss~constraint. For notation and labeling of edges we refer to figure \ref{tetra}. Also we will omit the index $v$ and instead of using an index $f,f'$ use a prime to denote objects associated to the face $f'$. 

The simplest example of a closed triangulation with a 3-valent vertex is the boundary of a tetrahedron. The dual is also a tetrahedral graph, and a basis of constraints is provided by choosing three faces around one dual vertex $v$. In this way we take into account the flatness constraints of three faces. The flatness constraint for the last, fourth face, is redundant due to the integral version of the Bianchi identity. Geometrically this choice of constraint basis means that we fix one triangulation vertex in 3D space and translate the other three triangulation vertices normal and tangential to the triangle described by $v$. There are still global rotations left, which lead to redundancies of constraints after implementation of the Gau\ss~constraints.

To find the Poisson brackets between the constraints we use the results developed in the previous section \ref{3}, but have to be careful to use the case $v=v'$ of the formula \eqref{pb3}. We find (see the Appendix \ref{app:calculations} for details) for the brackets between constraints in one and the same face $f$,
\be
\begin{aligned}
\{D_1,D_2\}&= h^0_f(D_1+D_2) \\
\{H,D_1\}&= - h^0_f H  + \tfrac{1}{2} \sqrt{\det{Q}}\, Q^{2\alpha}D_\alpha \, (Q^{1\beta}D_\beta - Q^{2\beta} D_\beta) \\
\{H,D_2\}&= h^0_f H -\tfrac{1}{2} \sqrt{\det{Q}}\, Q^{1\alpha}D_\alpha \, (Q^{1\beta}D_\beta - Q^{2\beta} D_\beta) \\
\{H,H\}&=0
\end{aligned}
\ee
where the sum in $\alpha,\beta$ is over $\alpha,\beta=1,2$. Also $\det{Q}=\det{Q_{\alpha\beta}}$ with $Q_{\alpha\beta}$ the inverse to $Q^{\alpha\beta}$.

For the constraints involving two faces $f,f'$, but one and the same three--valent vertex $v=v'$,  note that the normals coincide modulo Gau\ss~constraints $n=n'$. We also have $\det{Q}=\det{Q'}$ and
\ba
(Q')^{22}= Q^{11}+Q^{22}-2Q^{12}\, ,\q     (Q')^{23}=Q^{11}-Q^{12} \, ,\q  (Q')^{33}=Q^{11} \, .
\ea

The Poisson brackets between diffeomorphism constraints involving two faces $f,f'$ are
\be
\begin{aligned}
\{D_1,D_2'\} &= -h^0(D'_2 + D'_3) -\tfrac{1}{2} \sqrt{\det Q} \left[ Q^{12}(D_1H'+H(D'_2+D'_3)) +Q^{22}(D_2 H'- H D'_2)\right] \\
\{D_1,D'_3\} &= 0 \\
\{D_2,D'_2\} &= (h')^0 D_2+h^0D'_2.
\end{aligned}
\ee

A Hamiltonian and a diffeomorphism constraint give
\be
\begin{aligned}
\{H,D'_2\}&= h^0 H' - \tfrac{1}{2} \sqrt{\det Q} \bigl( Q^{1\alpha}D_\alpha\bigr) \bigl( (Q')^{2\gamma} D'_\gamma - (Q')^{3\gamma} D'_\gamma\bigr)\\
\{H,D'_3\}&=\tfrac{1}{2} \sqrt{\det Q}\ \bigl(Q^{2\alpha}\,D_\alpha\bigr) \ \bigl((Q')^{2\gamma}D'_\gamma\bigr)
\end{aligned}
\ee
with $\alpha=1,2$ and $\gamma=2,3$. Notice that the part linear in the constraints in $\{H, D_3'\}$ vanishes. Finally we obtain for the brackets of two Hamiltonians from two adjacent faces
\ba
\{H,H'\}&=& -(h')^0\, Q^{2\alpha}D_\alpha-h^0\, (Q')^{2\gamma}D'_\gamma  \, \q\q  \q .
\ea
The linear part of the Poisson bracket relations mirrors the continuum algebra, i.e. two diffeomorphism constraints give diffeomorphism constraints, a Hamiltonian and a diffeomorphism give a diffeomorphism whereas two Hamiltonians give diffeomorphism constraints. Also the last Poisson bracket algebra relation involves structure functions in the form of the inverse two--metric. This is the general form of the Poisson algebra, as expected from the geometric considerations of the Section \ref{5} and as we will see in arbitrary dimension in the Section \ref{simplex}. This form holds due to considering constraints around a three--valent vertex, where the splitting into Hamiltonian and diffeomorphism constraints is with respect to the same normal $n$. When the constraints involve different normals $n \neq \tilde n$, additional terms in the constraint algebra will appear. This is the purpose of the Section \ref{7}.


\section{Constraints at one face}\label{7}

\begin{figure}
\center
\subfigure[This is the face $f$ with vertices $v, v'$. To transport a vector from one to the other vertex, one uses the holonomies following the orientation of $f$.]{\includegraphics[scale=.6]{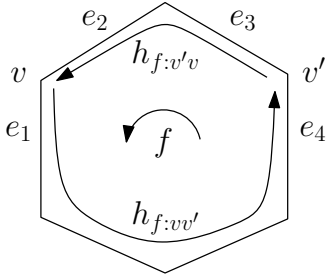} \label{fig:face} }
\hspace{3cm}
\subfigure[This is a gluing of triangles around a vertex, with the local frames associated to two different triangles displayed. The edge vectors $E_{1,2}$ and $E_{3,4}$ do not span the same plane, and the normals $n_v, n_{v'}$ do not coincide.]{\includegraphics[scale=.6]{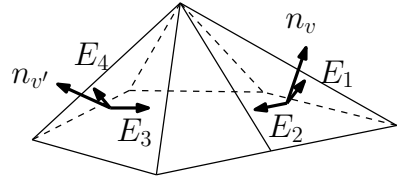} \label{fig:TriangulationFace} }
\caption{On the left is the face (here of degree 6) with the vertices $v$ and $v'$, and on the right the piece of triangulation (with 6 triangles) it is dual to. \label{fig:AroundFace}}
\end{figure}

We now consider the algebra of constraints on different vertices belonging to a single face (hence we will often drop the face label in the notations), as in the Figure \ref{fig:AroundFace}. At the vertex $v$, we have the local basis $(E_{\alpha v}, n_v)$ where $\alpha=1,2$ which is used to define the metric $Q_v$ and the constraints $D_\alpha^{v} = E_{\alpha v}^i h^i_{fv}$, $H^v=n^i h^i_{fv}$. At the vertex $v'$, we denote the edge vectors $E_{\alpha v'}$, the normal $n'$, the metric $Q_{v'}$ and the constraints $D^{v'}_\alpha, H^{v'}$.

Because we consider the brackets between constraints based on different vertices, their result is expressed in terms of quantities which have to be transported from $v$ to $v'$ (or the other way around). This involves the holonomy $h_{f:vv'}$ following the orientation of the face. We will denote $E_{\alpha v}(v') = \Ad_{h_{f:vv'}}(E_{\alpha v})$, $n_v(v') = \Ad_{h_{f:vv'}}(n_v)$. To go from $v'$ to $v$, this is the holonomy $h_{f:v'v}$ (following the orientation of the face, see the Figure \ref{fig:face}), and by definition $E_{\alpha v'}(v) = \Ad_{h_{f:v'v}}(E_{\alpha v'})$, $n_{v'}(v) = \Ad_{h_{f:v'v}}(n_{v'})$.

We find
\begin{multline}
\left\{ D_\alpha^v, D_\beta^{v'} \right\} = o_{f\alpha}\, h^0_{f}\, \bigl(E_{\beta v'}(v) \cdot E_{\lambda v}\bigr)\, Q^{\lambda \sigma}_v D_\sigma^{v} - o_{f\beta}\, h^0_{f}\, \bigl(E_{\alpha v}(v') \cdot E_{\lambda v'}\bigr)\, Q^{\lambda \sigma}_{v'} D_\sigma^{v'}\\
+ o_{f\alpha}\, h^0_{f}\,\bigl( n_v \cdot E_{\beta v'}(v)\bigr)\, H^v - o_{f\beta}\, h^0_{f}\,\bigl( n_{v'} \cdot E_{\alpha v}(v')\bigr)\, H^{v'}.
\end{multline}
The presence of $H^{v}$ and $H^{v'}$ in this bracket is due to the fact that the diffeomorphim constraints $D^v$ and $D^{v'}$ generate deformations in different planes (because the edge vectors $(E_{\alpha v})$ and $(E_{\alpha v'})$ do not span the same spatial slice, as emphasized in the Figure \ref{fig:TriangulationFace}). If the planes coincide, the prefactors in front of the Hamiltonians would vanish.

The bracket between a Hamiltonian and a diffeomorphism constraint gives
\begin{multline}
\left\{ H^v, D_\beta^{v'}\right\} =  - o_{f\beta}\,h^0_{f}\,\bigl(n_{v'}\cdot n_v(v')\bigr)\,H^{v'} + o_{f\alpha} \frac1{2\sqrt{\det Q}} \epsilon^{\lambda\sigma} \bigl(E_{\sigma v} \cdot E_{\beta v'}(v)\bigr) D_\lambda^v\,Q_v^{\alpha\gamma} D_\gamma^v\\
- o_{f\beta} h^0_{f} \bigl(n_v(v')\cdot E_{\beta v'}\bigr) Q_{v'}^{\beta \alpha} D_\alpha^{v'} - o_{f\alpha} h^0_{fv} \bigl(n_v\cdot E_{\beta v'}(v)\bigr) Q_v^{\alpha \gamma} D_\gamma^{v}.
\end{multline}
Among the linear part, the Hamiltonian is expected. The presence of the diffeomorphism contributions is again due to the fact that the local basis at $v$ and $v'$ do not coincide, since those terms come with the scalar product between the normal and the edge vectors (transported to the appropriate vertex).

In contrast, the linear part of the bracket between two Hamiltonians only generate diffeomorphism constraints,
\begin{multline}
\left\{ H^v, H^{v'}\right\} = o_{f\alpha} h^0_{f} \bigl(n_{v'}\cdot n_v(v')\bigr) Q_{v'}^{\alpha \beta} D_{\beta}^{v'} - o_{f\alpha} h^0_{f} \bigl(n_{v}\cdot n_{v'}(v)\bigr) Q_{v}^{\alpha \beta} D_{\beta}^{v}\\
+\frac12 o_{f\alpha} Q_{v'}^{\alpha \beta}D_\beta^{v'} \bigl((n_{v'}\times n_v(v'))\cdot E_{\gamma v'}\bigr) Q_{v'}^{\gamma\delta} D_\delta^{v'} - \frac12 o_{f\alpha} Q_{v}^{\alpha \beta}D_\beta^{v} \bigl((n_{v}\times n_{v'}(v))\cdot E_{\gamma v}\bigr) Q_{v}^{\gamma\delta} D_\delta^{v}.
\end{multline}

Therefore the linear part of the algebra mirrors quite well the continuum algebra. In fact, the differences can be explained in simple geometric terms. The local bases at $v$ and $v'$ are different so that a spatial (or normal) deformation at $v$ is typically not purely spatial (nor normal) anymore when transported to $v'$, but still decomposes onto $D^{v'}_\alpha$ and $H^{v'}$. (Notice that the transportation plays no role, since one can always gauge fix, say, $h_{f:vv'}=\mathbb{I}$.) This can be understood from the view of the triangulation dual to the face, where it is clear that what is tangential to the triangle dual to $v$ is not purely tangential to the triangle dual to $v'$.

\section{Geometric interpretation and constraints in Gau\ss~ reduced phase space}\label{8}

Holonomies and fluxes are the variables inherited from loop quantum gravity on a single graph. Since 3D gravity deals with flat (Euclidean) geometry, we should be able to re-write the constraints and their algebra in terms of rotation invariant, geometric quantities, like lengths and angles, obtained from the holonomies and the fluxes by performing a reduction with respect to the Gau\ss~ constraints \cite{waelbroeck}. We will later use these variables to generalize the constraint algebra (for the boundary of a simplex) to arbitrary dimensions.

To perform the phase space reduction with respect to the Gau\ss~constraints we need a set of rotation invariant variables. They are given by the extrinsic dihedral angles (a discretized form of the extrinsic curvature), which are associated to the edges $e$, and the lengths $l_e$ of the edges $e^*$.

\begin{figure}[hbt]
\begin{center}
\psfrag{1}{$e_1$}
\psfrag{2}{$e_2$}
\psfrag{3}{$e_3$}
\psfrag{4}{$e_4$}
\psfrag{5}{$e_5$}
\psfrag{6}{$e_6$}
\psfrag{v}{$v$}
\psfrag{f}{$f$}
\psfrag{fp}{$f'$}
 \includegraphics[scale=0.5]{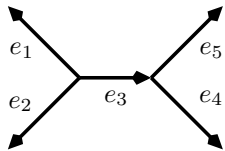}
 \caption{Definition of the dihedral angle.} \label{dihedralf}
 \end{center}
\end{figure}

The dihedral angle is defined as the angle between the normals of two neighbouring triangles. Here we need however to parallel transport one of the normals, so that we can compute the inner product in the same coordinate system. To avoid factor ordering ambiguities (for quantization) we choose for the definition of the normals the two edges which are not shared with the other triangle, see figure \ref{dihedralf}. Thus for the extrinsic dihedral angle at the edge $e_3$ in figure \ref{dihedralf}
\ba\label{dihedral}
\cos \Theta_3 &=& \frac{(E_4 \times E_5) \,\,\cdot \,\, g_3  ( E_1 \times E_2) g_3^{-1} }{|E_4 \times E_5| \,\, |E_1 \times E_2| }\,=\, \frac{N'\cdot (g_3 N g_3^{-1})}{|N'|\,|N|}
\ea
with $N=E_1\times E_2$ and $N'=E_4\times E_5$.
The length $l_e$ of the dual edge $e^*$ is given as $l_e=\sqrt{E^kE_k}$.

 From the Poisson brackets of the fluxes and holonomies (\ref{fluxhol}) one can find \cite{waelbroeck} the Poisson bracket relations between these variables as
\ba\label{lengththeta}
\{l_{e},l_{e'}\}=0 \q ,\q\q \{\Theta_e, \Theta_{e'}\}=0 \q ,\q\q \{ l_{e}, \Theta_{e'}\}=-\delta_{ee'} \q .
\ea
For instance for the last Poisson bracket in the case $e'=e$ consider the set up of figure \ref{dihedralf}. One finds that
\ba
\{ E_3\cdot E_3, N'\cdot (g_3 N g_3^{-1}) \}&=& 2 \, (g_3 (E_3\times N)g_3^{-1} ) \cdot N' \,=\,   2\, |N| \, |N'| \, |E_3| \,\sin \Theta_3
\ea
from which $\{l_{3}, \Theta_3\}=-1$ follows. 

We will  express the Hamiltonian and diffeomorphism constraints in scalar variables. This allows to bring the constraints into their simplest form (for the tetrahedron), which can easily be generalized to higher dimensions.

Consider the Hamiltoninian $H^{fv}$ where $v$ is a vertex with three outgoing edges $e_1,e_2,e_3$ and $f$ is a face bounded by the edges $e_1,e_6,e_2^{-1}$, see figure \ref{tetra}. This Hamiltonian is given by
\ba
H^{fv}&=&\frac{ (E_1\times E_2)^k}{|E_1\times E_2|} (-2) \tr(h_{fv}T^k) \nn\\
&=& \frac{1}{|E_1\times E_2|} (-2) \tr( [E_1,E_2] g_2^{-1} g_6 g_1) \nn\\
&=&\frac{1}{|E_1\times E_2|} (-2) \left(  \tr(E_{\bar{1}} (g_1g_2^{-1} E_{\bar{2}} g_6 ) \,-\, \tr( E_2 E_1 g_2^{-1} g_6 g_1)\right) \q .
\ea
We will now use the flatness constraint for the face $f$ itself in order to replace $g_6$ with $g_6=g_2 g_1^{-1}$. This will lead to a new constraint $(H')^{fv}$, which is a combination of the constraints for the face $f$. It is however straightforward to check, that this constraint has the same action on the fluxes $E_1,E_2,E_6$ (on the constraint hypersurface) as the constraint $H^{fv}$. In the following we will not distinguish between (flatness) constraints leading to same geometric action on the fluxes\footnote{The tetrahedron has zero physical degrees of freedom, that is the constraints determine completely the dihedral angles as functions of the lengths. Thus on the constraint hypersurface the action of two constraints on the dihedral angles agrees if the action agrees for the length variables, or equivalently the fluxes.} and drop the prime in $(H')^{fv}$. This will not affect the terms linear in the constraints in the Poisson algebra. Thus we redefine
\ba\label{hnew}
H^{fv}:=\frac{1}{|E_1\times E_2|} \left(   (g_6 E_{\bar{1}} g_6^{-1})\cdot E_{\bar{2}} - E_1\cdot E_2\right) \q .
\ea
The corresponding density two version (without the denominator) has been quantized in \cite{valentin1}.


Now the exterior dihedral angle $\Theta_6$ can be defined as follows
\ba
\cos \Theta_6 &=& \frac {N'' \cdot (g_6 N' g_6^{-1}) }{|N'| \, |N''|}  \q \text{with} \q\q
N'= E_6 \times E_{\bar{1}}   \, ,\q\q
N''= E_{\bar{2}} \times  E_{\bar{6}}
\ea
which on the Gau\ss~constraint surface is equivalent to  the definition (\ref{dihedral}).
Expressing the normals $N',N''$ through the fluxes one can rewrite this exterior angle as
\ba
\cos \Theta_6 &=& -\frac{\cos \rho - \cos \alpha_{16} \cos\alpha_{26}}{\sin \alpha_{16} \sin \alpha_{26}}
\ea
with $\alpha_{16}, \alpha_{26}$ the (interior) 2D angles at $f^*$ spanned between the dual edges $e_1^*,e_6^*$ and $e_6^*,e_2^*$ respectively. Here the angle $\rho$ is defined as
\ba
\cos \rho&=&-\frac{E_{\bar{2}}\cdot (g_6 E_{\bar{1}} g_6^{-1})} {|E_1|  \, |E_2|} \q .
\ea
Compare this with the definition of the (internal) 2D angle $\alpha_{12}$ between $e_1^*$ and $e_2^*$
\ba
\cos \alpha_{12}&=&-\frac{E_{2}\cdot E_1 } {|E_1| \, |E_2|} \q .
\ea
In the case that the 2D surface is embedded in 3D flat space the parallel transport of a vector around the dual vertex $f^*$ is trivial and we should have
\ba
\cos\alpha_{12}&=&\cos \rho \q .
\ea
Indeed this defines the `flat' (exterior) dihedral angle as a function of the 2D angles (and hence the lengths of the six edges making up the three triangles meeting at $f^*$)
\ba
\cos \Theta_6^{flat} (\alpha_{12},\alpha_{16},\alpha_{26}) &=& -\frac{\cos \alpha_{12} - \cos \alpha_{16} \cos\alpha_{26}}{\sin \alpha_{16} \sin \alpha_{26}} \q .
\ea
This relation generalizes to higher dimensions \cite{areaangle}, e.g. holds also between the 4D (exterior) dihedral angles and 3D (interior) dihedral angles of a 4--simplex and even for homogeneously curved simplices \cite{newangle}.

Now the Hamiltonian (\ref{hnew}) is given as
\ba\label{hnew2}
H^{fv} &=& \frac{|E_1|\,|E_2|}{|E_1\times E_2|}\,(\cos\alpha_{12} \,-\,\cos \rho) \nn\\
&=&\frac{\sin \alpha_{16} \sin \alpha_{26}}{\sin \alpha_{12}} \left( \cos \Theta_6 -\cos \Theta_6^{flat}\right) \q .
\ea

The term in brackets
\ba\label{quadC}
C_\beta &=& \cos \Theta_\beta - \cos \Theta^{flat}_\beta
\ea
defines the simplest `quadratic form' (or rather cosine form) of the constraints for a tetrahedron (with $\beta=1,\ldots,6$ denoting the edges) with the clear geometric meaning to impose the dihedral angles as those from a geometric tetrahedron. This form allows for both positive and negative dihedral angles according to considering the outside or the inside of the tetrahedron as the 2D surface. It appeared in \cite{dittrichryan,dittrichreview,valentin4D} and as we will see later this form can easily be generalized to higher dimensions.  The `linearized form' (there are two sectors $\pm$ corresponding to positive or negative orientation of the simplex)
\ba\label{linC}
C^{\pm}_\beta &=& \Theta_\beta \mp  \Theta^{flat}_\beta
\ea
defines Abelian constraints. This follows from the Schl\"afli identity for the variations of the dihedral angles in a tetrahedron
\ba
\sum_{\beta } l_\beta \, \delta \Theta_\beta &=& 0
\ea
where $\beta$ labels the edges in a tetrahedron. The Schl\"afli identity  ensures that the dihedral angles are generated as derivatives with respect to the lengths from the Regge action $\sum_\beta l_\beta \Theta_\beta$ for one tetrahedron. Thus the dihedral angles are indeed the conjugated momenta to the length variables. As the dihedral angles can be obtained from a generating function, namely the Regge action, we have
\ba
\frac{\partial}{\partial l_\beta} \Theta_\alpha &=&  \frac{\partial}{\partial l_\alpha} \Theta_\beta
\ea
which shows that the constraints (\ref{linC}) are Abelian.


A  form of the constraint (\ref{hnew2}) linear in the dihedral angles, that agrees with (\ref{hnew2}) in its action on the length variables on the part of the constraint hypersurface  describing the positive orientation solution of the tetrahedron, is given by
\ba
H &=& \frac{-\sin \Theta_6^{flat}   \sin \alpha_{16} \sin \alpha_{26} }{\sin \alpha_{12}} \left( \Theta_6  - \Theta_6^{flat}\right) \nn\\
&=&- \frac{3}{l_6}  \frac{V}{V_{{v^*}}}\left( \Theta_6  - \Theta_6^{flat}\right)
\ea
where $V$ is the volume of the tetrahedron with edge lengths $l_1,\ldots, l_6$ and $V_{v^*}$ is the volume of the triangle $v^*$ (with edge lengths $l_1,l_2,l_3$).  Here we fix one orientation of the tetrahedron (the boundary is given by the outside of the tetrahedron). Thus this constraint is not fully equivalent to the version (\ref{hnew2}), which allows for both orientations. Note that we can switch between the linear (in the dihedral angles) and cosine form of the constraints by making the exchange $-\sin \Theta^{flat} \, \Theta \leftrightarrow \cos \Theta$.

Similarly we can find the Hamiltonian for the face $f'$, which is bounded by the edges $e_2,e_4,e_3^{-1}$, so that
\ba\label{hams}
H&=&- \frac{3}{l_6}  \frac{V}{V_{{v^*}}}\left( \Theta_6  - \Theta_6^{flat}\right) \nn\\
H'&=&-  \frac{3}{l_4}  \frac{V}{V_{{v^*}}}\left( \Theta_4  - \Theta_4^{flat}\right)  \q . 
\ea

The  diffeomorphism constraints can also be deduced from their action on the length variables. One finds
\ba\label{diffs}
D_1 &=& -l_1 (\Theta_1-\Theta_1^{flat}) -   l_2  \cos\alpha_{12}  (\Theta_2-\Theta_2^{flat}) -l_6\cos\alpha_{16}(\Theta_6  - \Theta_6^{flat}) \nn\\
D_2 &=& l_2  (\Theta_2-\Theta_2^{flat}) + l_1 \cos\alpha_{12}  (\Theta_1-\Theta_1^{flat}) + l_6 \cos\alpha_{26}(\Theta_6  - \Theta_6^{flat}) \nn\\
D'_2 &=&     -l_2    (\Theta_2-\Theta_2^{flat}) - l_3 \cos\alpha_{23}  (\Theta_3-\Theta_3^{flat}) - l_4 \cos\alpha_{24}(\Theta_4  - \Theta_4^{flat})        \nn\\
D'_3&=& l_3 (\Theta_3-\Theta_3^{flat})+ l_2 \cos\alpha_{23}    (\Theta_2-\Theta_2^{flat}) +   l_4 \cos\alpha_{34}(\Theta_4  - \Theta_4^{flat}) \q .
\ea

The constraints (\ref{hams}) and (\ref{diffs}) form a first class algebra. As we now have a version of the constraints linear in the momentum variables, this algebra will only include linear terms in the constraints. These reproduce the linear part of the constraint algebra found in section \ref{6}.


\section{Thiemann's rewriting} \label{9}

The work \cite{valentin1} used a density 2 version of the Hamiltonian constraint (\ref{hnew}). This has the advantage that no factor ambiguities arise and that further more one does not need to worry of how to define the inverse of $|E_1 \times E_2|$, which can be understood as the  volume density of the spatial hypersurface.  However to draw some lessons for the 4D theory one might want to address the factor ordering ambiguity as well as work with the density one version of the constraint, which is argued \cite{qsd} to be the  valid version allowing for a  continuum limit. Furthermore these versions might differ in their action on degenerate configurations, which might have repercussions on the choice of boundary conditions for the physical solutions in the quantum theory, in particular the boundary condition for zero size hypersurfaces (the `big bang' or `big crunch'). Thus the density two version of the Hamiltonian leads to a physical wave function describing both orientations of the spherical universe at once  \cite{valentin1}.

The problem of how to divide by the inverse volume in the quantum theory was solved in 4D \cite{qsd} and in 3D \cite{qsd3D} by the so--called Thiemann trick, which basically uses that
\ba
\frac{1}{\sqrt{V}} = 2 \frac{ \bd }{\bd V} \sqrt{V}
\ea
and expresses the derivative via Poisson brackets.  These identities where derived for the continuum theory. We want to point out here, that such an identity can equally well be derived directly for the discrete geometry.

Indeed, in the context of our discrete geometry, we can express the normalized normal of a triangle, spanned by dual edges $e_1^*,e_2^*$ (with $e_1,e_2$ outgoing edges at a vertex $v$ and the holonomy $h_{fv}$ starting with $e_1$) as\footnote{Note that it is important to use for the normal $N^i= {\epsilon^i}_{mn}E^m_1 E^n_2 $ and not to replace one of the fluxes $E_1,E_2$ by $E_3$ via the Gau\ss~constraints. This would change the result of the Poisson bracket, which involve $g_1,g_2$ and hence objects not invariant under rotations generated by the Gau\ss~constraints. }
\ba
n^i=\frac{1}{\sqrt{N\cdot N}} N^i &=& 4\, {\epsilon^i}_{mn} R^m_1 R^n_2 \q\q \text{where}  \nn\\
R_1^m&=&-2 \,\tr \left( \,g_1^{-1}\, \{(N\cdot N)^{\frac{1}{4}}, g_1 \} \,T^m \right)  \q , \nn\\
R_2^m&=&-2 \tr\, \left(g_2^{-1}\, \{(N\cdot N)^{\frac{1}{4}}, g_2 \} \,T^m \right) \q\q \text{and}   \nn\\
N^i&=& {\epsilon^i}_{mn}E^m_1 E^n_2  \q .
\ea
This gives for the contraction of $F_{f,v}$ (where the face $f$ is defined by the edges $e_1,e_2$) with the normal $-n^i$
\ba\label{symham}
H^{fv}=-\sum_k F_{f,v}^k n^k &=& 8\, \tr\left(F_{f,v} [R_1,R_2]  \right)
\ea
where the square brackets are the Lie algebra brackets $[R_1,R_2] =R_1R_2 -R_2R_1$ and
\ba
R_\alpha=\sum_m R_\alpha^m T^m &=& g_\alpha^{-1}  \, \{(N\cdot N)^{\frac{1}{4}}, g_\alpha \}
\ea
with $\alpha=1,2$. This results in an expression similar to the one used in \cite{qsd3D} for the quantum Hamiltonian. The difference is that here we do not necessarily average over the faces adjacent to $v$. For future work it will be interesting to compare the different factor orderings of (\ref{symham}) to the (density two) Hamiltonian in \cite{valentin1}.

\section{4D gravity: flat sector} \label{sec:4Dflat}

The hypersurface deformation algebra of constraints can in the case of 3D discrete gravity be defined for arbitrary triangulated (or polygonated) 2--surfaces. It is a first class algebra -- as already follows from the fact that the constraints arise from combinations of the flatness constraints, which are Abelian.

However, 3D gravity is in a sense exceptional, as here discretization does not break the diffeomorphism symmetry of the theory -- it is preserved in the form of vertex translation symmetry \cite{louapre,dittrichreview}. This symmetry can also be translated into the quantum theory \cite{barrettcraneH,louapre,valentin1}. This already changes if we add a cosmological constant, here standard Regge calculus (with flat simplices) does break diffeomorphism symmetry \cite{bahrdittrich2}, in the sense that there are no transformations that leave the action invariant and act non--trivially on solutions. An alternative discretization employing homogeneously curved simplices \cite{bahrdittrich2,newangle} can however be constructed, which again does preserve the symmetries.

3D gravity is a topological theory and thus one might argue that diffeomorphism symmetry can only be preserved for these kind of theories. Indeed the other known examples, where diffeomorphism symmetry (or rather reparametrization invariance) is preserved are $(0+1)$ dimensional \cite{steinhaus1} and thus also topological. However also in more complicated theories diffeomorphism symmetry can be preserved, either by admitting non--local discretizations \cite{dittrichzako} or by using the concept of cylindrical consistency to allow for more complicated building blocks \cite{cylindrical}.

Although diffeomorphism symmetry in 4D gravity is generically broken, residual gauge symmetries still remain \cite{bahrdittrich1}. For vertices with adjacent triangles carrying curvature, the vertex translation symmetry is  broken, to an order quadratic in the deficit angle (which is proportional to the curvature). There are however flat solutions or even vertices which are embedded into a flat neighbourhood, for which vertex translation symmetry is extant.

A canonical formulation of 4D discrete gravity can be either obtained directly from Regge calculus \cite{bahrdittrich1,hoehn}, or via a Gau\ss~ constraint reduction from a loop quantum gravity like discretization, which starts with connection and bi--vector variables \cite{zapata4d, dittrichryan, dittrichryan23}.

In the analysis starting directly from Regge calculus one finds (Abelian) constraints for the linearized theory for every vertex, which are however changed into proper equations of motion if the non--linear order is taken into account \cite{hoehn}. Again there are special configurations, for instance a four--valent vertex in the 3--dimensional triangulated hypersurface, for which the constraints survive to any order. The reason is that such a vertex leads to a flat neighbourhood in the 4--dimensional solution.

Indeed as was first pointed out in \cite{dittrichryan}, there is a family of triangulations of the boundary of the 4--sphere, namely boundaries of so--called stacked spheres, which lead to flat bulk solutions. For these triangulations we can define first class constraints. The simplest example is the boundary of the 4--simplex to which we will restrict in the following section. (These considerations easily generalize to constraints around a four--valent vertex in any triangulation.)

The work \cite{zapata4d, dittrichryan} starts, as in 3D, with a discretization using holonomies and bi--vectors. This is based on a discretization from the Plebanski action, which is employed in spin foams. As is well known, the 4D case is much more challenging than 3D, due to the appearance of primary and secondary simplicity constraints which are mostly (in the discrete theory) second class constraints. As was first pointed out in \cite{dittrichryan} the reduction by these simplicity constraints can be performed in two stages. The first stage reduces to a phase space analogous to loop quantum gravity restricted to the dual graph of the triangulation. However, this phase space is strictly bigger than the phase space corresponding to Regge calculus. The configurations described by this phase space were later coined twisted geometries \cite{twisted}. The second stage involves the so--called gluing constraints \cite{areaangle}, which are partially second class and partially first class. This last first class part are known as area constraints \cite{williamsarea} -- this set is however empty for the boundary of a simplex. This reduction recovers the symplectic structure found directly for Regge calculus \cite{hoehn} and allows to express the constraints for the 4--simplex in the same simple  form as for the tetrahedron (\ref{quadC},\ref{linC}). Indeed the constraints are of the same form.

Because of the complications which arise in the reduction from bi--vectors and holonomies to scalar variables, we will consider only scalar variables in this work, and leave the investigation of the larger phase space for loop quantum gravity  for future work. Also, we want to point out that the Plebanski action agrees with the (topological) BF action to which the simplicity constraints are added. BF theory leads again to the flatness constraints, and the corresponding quantization on a simplex has been considered in \cite{valentin4D}.

\section{The constraints for the boundary of a simplex}\label{simplex}

The following considerations will hold for the boundary of a $d$--dimensional simplex $\sigma$, with $d\geq 3$.
For such a simplex it is convenient to introduce the following notation: We will label the vertices of the simplex with $i=0,\ldots, d$ and denote by $\sigma ,\sigma(i),\sigma(ij)$ etc.\ the $d$--simplex itself, the subsimplex  of $\sigma$ which does not include the vertex $i$ and the subsimplex of $\sigma$ which does not include the vertices $i,j$ respectively. Correspondingly $V,V(i),V(ij)$ will denote the volumina of these various simplices. Also $\theta(ij)$ will be the {\it internal} dihedral angle between the subsimplices $\sigma(i)$ and $\sigma(j)$, hence $\theta(ij)$ is associated to the $(d-2)$-simplex $\sigma(ij)$.

The phase space variables for the boundary of a $d$--simplex can be taken to be the volumina\footnote{There are  $\tfrac{1}{2}d(d+1)$ of these volumina  and $\tfrac{1}{2}d(d+1)$ lengths variables for a $d$--simplex. The volumina $V(ij)$ can be uniquely determined form the length variables, for $d\geq 4$ there is however a discrete ambiguity in the transition from the volumina to length variables. Singularities in this map appear for configurations involving orthogonal angles. Away from these configurations the map and its inverse can however be defined locally. }  $V(ij),\, j>i$ and the  {\it internal} dihedral angles $\theta(kl),\,l>k$:
\ba
\{V(ij),\theta(kl)\} &=& \delta_{(ij),(kl)} \q .
\ea

The constraints express that the dihedral angles are actually fixed as functions of the geometry:
\ba\label{4dc1}
C_{(ij)}=\theta_{ij} \,-\, \theta^{flat}_{ij}
\ea
where $\theta^{flat}_{ij}$ is the geometric internal dihedral angle determined from the volumina of the simplex. These constraints are Abelian, again due to the Schl\"afli identity. Indeed in form they coincide with (\ref{linC}) for 3D gravity. More generally the same constraints and phase space variables can be defined for any $d$--simplex (with $d\geq 3$).

Also the geometric action of the constraints (\ref{4dc1}) is clear: it changes the geometry of the simplex such, that only $V(ij)$ is altered. As we have one constraint for every subsimplex $\sigma(ij)$, whose volumina parametrize (locally) the geometry, we can change this geometry in an arbitrary way by the action of the constraints. Therefore with the appropriate linear combination of constraints we can reproduce the change of an arbitrary vertex translation. The algebra will remain first class, as we just take combinations of constraints.

Thus we have to find the variation of the $V(ij)$  under the various deformations of the geometry induced by vertex translations. To this end we find it convenient to introduce affine coordinates and the affine metric for the simplex, which are explained in appendix \ref{A}. The dihedral angles and the various volumina are given by components of the inverse affine metric.

As for the tetrahedron in the Section \ref{6}, to obtain a set of constraints which allows for all possible deformations of the geometry, we can fix the vertex $0$. We will consider translations of the other $d$ vertices $k=1,\dotsc,d$.
\begin{itemize} 
 \item We define $d$ Hamiltonians $H(k)$, one for each vertex $k=1,\dotsc,d$, by requiring that $H(k)$ translates the vertex $k$ along the outward pointing unit normal to $\sigma(0)$.
 \item The diffeomorphism constraint denoted by $D(kl)$ will translate the vertex $k$ along a vector that is tangential to $\sigma(0)$ and normal to $\sigma(0l)$. The reason for this choice is that it corresponds to the geometric action of the constraints used in 4D loop quantum gravity which in form (for the Barbero--Immirzi parameter that leads to a self dual connection)  are the same as the 3D constraints (\ref{h7},\ref{h8}).
\end{itemize}

The geometry of the flat simplex is conveniently described by its affine metric (background material can be found in the Appendix \ref{A}). The change of the affine metric $\tilde g_{ij}$ induced by a translation of the vertex $k$ by a vector $v$ of affine components $v^m$ is given by
\ba\label{gchange}
\delta_{k,v} (\tilde g_{ij}) &=& \tilde\delta^k_i  v^l \tilde g_{lj} + \tilde g_{il}v^l \tilde \delta^k_j \q .
\ea
Here $N(k)_i=\tilde \delta^k_i$ is the (inward pointing) normal to the subsimplex $\sigma(i)$ with norm
\ba
N(k)_i \tilde g^{ij} N(k)_j &=& \tilde g^{kk} \;=\;  \frac{1}{d^2} \frac{V(k)^2}{V^2}
\ea
where $d$ is the dimension of the simplex $\sigma$. It appears in (\ref{gchange}) as $N(k)_i (e_{mn})^i=0$ for edge vectors $e_{mn}$, with $m,n \neq k$.  This ensures that the lengths of such vectors is not changed under the displacement of the vertex $k$. On the other hand we have for edge vectors $e_{mk}$
\ba
\delta_{k,v}(l_{mk}^2)\;=\; e_{mk}^i \delta_{k,v} (\tilde g_{ij}) e_{mk}^j &=& 2 v^l  \tilde g_{lj} e_{mk}^j
\ea
as one expects for the change of the length square under a displacement of the vertex $k$ by a vector $v$.

Thus for the Hamiltonian $H(k)$ we have to use a deformation vector $v=-\hat N(0)$ where the hat $\hat{~}$  signifies normalization to one:
\ba\label{hamd}
\delta_{H(k)} (\tilde g_{ij})&=&-  d\frac{V}{V(0)} \left(  \tilde \delta^k_i \tilde \delta^0_j + \tilde \delta^0_i \tilde \delta^k_j\right) \;=\; -      \frac{1}{\sqrt{\tilde g^{00}}}   \left(  \tilde \delta^k_i \tilde \delta^0_j + \tilde \delta^0_i \tilde \delta^k_j\right)    \q .
\ea

To define the diffeomorphism constraints we need the vectors tangential to $\sigma(0)$ but normal to the subsimplex $\sigma(0l)$. These can be found via the induced metric for the subsimplex $\sigma(0)$
\ba\label{inducedmetric}
\tilde h(0)_{ij} &=& \tilde g_{ij} - \hat N(0)_i \hat N(0)_j \q .
\ea
This gives the following projector onto the simplex $\sigma(0)$ (indices are still raised and lowered with the full metric)
\ba
\tilde h(0)^i_j &=& \tilde \delta^i_j - \frac{\tilde\delta^0_j\, \tilde g^{0i}}{\tilde g^{00}}  \q ,
\ea
so that we can define the normal $N(l|0)$, tangential to $\sigma(0)$ and orthogonal to $\sigma(0l)$ as
\ba
N(l|0)_m&=& \tilde h(0)^l_m
\ea
with norm
\ba \label{h^ll}
  N(l|0)_k N(l|0)^k&=& h(0)^{ll} \;=\; \tilde g^{ll}-\frac{\tilde g^{l0}\tilde g^{l0}}{\tilde g^{00}}  \;=\; \frac{1}{(d-1)^2} \frac{ V(0l)^2}{V(0)^2} \q .\
\ea

The  diffeomorphism constraints $D(kl)$ translate the vertex $k$ by
\ba
 v=- \check N(l|0)&=&- (d-1) V(0) \,\, N(l|0) \q ,
\ea
thus the induced change on the affine metric is given by
\ba\label{diffeod}
\delta_{D(kl)}( \tilde g_{ij}) &=& - (d-1) V(0) \left( \tilde\delta^k_i \tilde\delta^l_j  +  \tilde\delta^k_j  \tilde\delta^l_i    -\frac{\tilde g^{0l}}{\tilde g^{00} }(\tilde\delta^k_i \tilde\delta^0_j  +\tilde \delta^k_j \tilde\delta^0_i) \right) \q .
\ea

The linear constraint $C_{ij}$ in  (\ref{4dc1}) generates changes in  the area $V(ij)$. To find the constraint corresponding to a deformation $\delta_{k,v}$ we therefore have to set
\ba
C_{k,v}:=\sum_{i<j} \delta_{k,v}(V(ij)) \,\, C_{(ij)} \q .
\ea

To find the change in the volumina $V(ij)$, note the following identity for the volume of a simplex (see the Appendix \ref{A})
\ba
\delta_{k,v} V&=& \frac{1}{2} V \tilde g^{mn} \,\,\delta_{k,v}(\tilde g_{mn}) \q .
\ea
This also generalizes to subsimplices, if one uses the induced metric to project the variation of the full affine metric:
\ba
\delta_{k,v} V(k)&=& \frac{1}{2} V(k) \tilde h(k)^{mn} \,\delta_{k,v}(\tilde g_{mn}) \nn\\
\delta_{k,v} V(kl)&=& \frac{1}{2} V(kl) \tilde h(kl)^{mn} \delta_{k,v}(\tilde g_{mn}) \q
\ea
where in analogy to (\ref{inducedmetric}) the induced metric on the simplex $\sigma(kl)$ is defined as
\ba
\tilde h(kl)_{mn} &=& \tilde h(k)_{mn} - \frac{\tilde h(k)^l_{m} \tilde h(k)^l_n}{h(k)^{ll}}  \q .
\ea

Thus
\ba
C_{k,v} &=& \frac{1}{2} \sum_{k<l,m,n}  V(kl)  \,\tilde h(kl)^{mn} \delta_{k,v} \, (\tilde g_{mn}) \,\, (\theta_{kl} -\theta^{flat}_{kl})
\ea
where for $\delta_{k,v} \, (\tilde g_{mn})$ we have to use (\ref{hamd}) for the Hamiltonian constraints and (\ref{diffeod}) for the diffeomorphism constraints.

\section{The simplex boundary deformation algebras} \label{sec:bsda}

In this way we obtain an explicit realization of the constraint algebra, however for finding the commutator of these constraints we do not need this representation. This is due to the linearity of the constraints in the momentum variables -- would we use some other representation of the constraints the strategy which we are going to use now will only give the part of the constraint algebra that is linear in the constraints, i.e. the part important for the flow on the constraint hypersurface.

To find the algebra of the constraints, we will consider the algebra of the deformations induced by these constraints \cite{teitelboim},
\ba\label{n1}
 \{ f , \{ C_{k',v'}, C_{k,v} \} \} &=& \{ \{ f, C_{k',v'}\},C_{k,v}\} - \{\{f, C_{k,v}\},C_{k'v'} \} \nn\\
 &=& \left[ \delta_{k,v} \circ \delta_{k',v'} -\delta_{k',v'}\circ \delta_{k,v}\right](f) \q.
\ea
Knowing the action of the combination of deformations in the last line of (\ref{n1}) on all metric elements $f=\tilde g_{mn}$ will allow us to deduce the Poisson brackets $\{ C_{k',v'}, C_{k,v} \} $. Applying two deformations to the affine metric we can write
\ba
 \delta_{k',v'}(\delta_{k,v} \tilde g_{ij})&=&\delta_{k'v'}\left(  \tilde \delta^k_i v_j +\tilde \delta^k_jv_i   \right) \nn\\
 &=&\tilde \delta^k_i \, \delta_{k'v'}(v_j) + \tilde \delta^k_j\, \delta_{k'v'}(v_i) \q.
\ea
where $v_i = v^j \tilde{g}_{ij}$. Thus we have to determine how the components $v_i$ corresponding to the Hamiltonian or diffeomorphism constraints change under a deformation $\delta$.

The easiest example is the unit normal to the simplex $\sigma(0)$
\ba\label{normald}
\delta\hat N(0)_j &=& \delta\left( \frac{1}{\sqrt{g^{00}}} \tilde \delta^0_j \right) \;\;=\; \;\frac{1}{2} \hat N(0)_j \frac{\tilde g^{0m}\tilde g^{0n}}{\tilde g^{00}}  \,\delta(\tilde g_{mn})
\ea
For the normal $\check{N}(l|0)_j$ one finds
\ba\label{subnormald}
\delta( \check{N}(l|0)_j ) &=& (d-1)\,\, \delta\left( V(0) \,\left( \tilde \delta^l_j -\frac{\tilde \delta^0_j \tilde g^{0l}}{\tilde g^{00} } \right) \right) \nn\\
&=& \check{N}(l|0)_j \, \frac{1}{2} \tilde h(0)^{mn} \delta(\tilde g_{mn}) + (d-1) V(0) \, \tilde \delta^0_j \frac{\tilde g^{0m}}{\tilde g^{00}}\, \tilde h(0)^{ln} \, \delta(\tilde g_{mn}) \q .
\ea

Let us consider the commutator between two Hamiltonian constraints $H(k)$ and $H(k')$.  For the change of $-\hat N(0)_j$ under the Hamiltonian constraint $H(k')$ we have to use (\ref{hamd}) in (\ref{normald}) and find
\ba\label{term1}
\delta_{H(k')} \bigl(-\hat N(0)_j\bigr) &=&\frac{\tilde g^{0k'}}{\sqrt{\tilde g^{00}}} \, \hat N(0)_j \q .
\ea
This defines a deformation vector that we use in (\ref{gchange}) to get
\ba
\delta_{H(k)} \circ \delta_{H(k')}(\tilde g_{ij}) &=&  \frac{\tilde g^{0k'}}{\tilde g^{00}}  \left( \tilde \delta^k_i \tilde \delta^0_j  \, +\,  \tilde \delta^k_j \tilde \delta^0_i \right)
\ea
and similarly
\ba\label{term2}
-\delta_{H(k')}\circ \delta_{H(k)}(\tilde g_{ij}) &=&-\frac{\tilde g^{0k}}{\tilde g^{00}}  \left( \tilde \delta^{k'}_i \tilde \delta^0_j  \, +\,  \tilde \delta^{k'}_j \tilde \delta^0_i \right) \q .
\ea

Comparing with (\ref{hamd}) the sum of this terms could be interpreted\footnote{Thus even the algebra involving only Hamiltonian constraints is closed. Using  density two Hamiltonians should just add terms proportional to Hamiltonian constraints.}
as proportional to a Hamiltonian deformation at vertex $k$ and a Hamiltonian deformation at vertex $k'$. The Hamiltonian and diffeomorphism deformation vectors we defined are however still overcomplete. Not considering deformations at the vertex $0$ fixes the global translations, we have however still the global rotations left. Thus this combination of Hamiltonian deformations can be rewritten as a combination of diffeomorphism deformations. To this end we have to add and subtract the appropriate term, so that
\ba
\left(\delta_{H(k)} \circ \delta_{H(k')}(\tilde g_{ij}) - \delta_{H(k')}\circ \delta_{H(k)}(\tilde g_{ij})\right)
&=&
-\tilde \delta^k_i \tilde \delta^{k'}_j - \tilde \delta^{k'}_i \tilde \delta^k_j + \tilde \delta^k_i \tilde \delta^{k'}_j + \tilde \delta^{k'}_i \tilde \delta^k_j   \nn\\&&
+ \frac{\tilde g^{0k'}}{\tilde g^{00}}  \left( \tilde \delta^k_i \tilde \delta^0_j  \, +\,  \tilde \delta^k_j \tilde \delta^0_i \right)
-\frac{\tilde g^{0k}}{\tilde g^{00}}  \left( \tilde \delta^{k'}_i \tilde \delta^0_j  \, +\,  \tilde \delta^{k'}_j \tilde \delta^0_i \right)  \nn\\
&=&
\frac{1}{(d-1) V(0)} \left( \delta_{D(kk')} - \delta_{D(k'k)} \right)(\tilde g_{ij})\q .
\ea
Thus we obtain the Poisson bracket between two Hamiltonians
\ba
\{H(k), H(k') \} &=& \frac{1}{(d-1)V(0)} \, \left( D(k'k)-D(kk')\right) \q .
\ea
For the Poisson brackets between two diffeomorphisms we have to follow the same strategy and in the end add and subtract the appropriate term to rewrite the result into a combination of diffeomorphism constraints again (otherwise Hamiltonian constraints are appearing). The Poisson brackets between a Hamiltonian and a diffeomorphism is the simplest case, as the normal $\hat N(0)_j$ does not change under a diffeomorphism. 
In summary we obtain the simplex boundary deformation algebra
\ba\label{sbda}
\{H(k), H(k') \} &=& \frac{1}{(d-1)V(0)} \, \left( D(k'k)-D(kk')\right)  \nn\\
\{D(kl),H(k')\} &=& (d-1) V(0) \,\tilde h(0)^{lk'} \,\, H(k) \nn\\
\{D(kl),D(k'l')\} &=& (d-1) V(0) \left(   \tilde h(0)^{kl} D(k'l') -\tilde h(0)^{k'l'}D(kl)+ \tilde h(0)^{ll'} \bigl( D(kk')-  D(k'k)\bigr) \right)\,. \nn\\
\ea
The somewhat complicated relations involving the diffeomorphism constraints are due to the choice\footnote{This is motivated by the geometric interpretation of the flux variables in the 4D case.} we have made to use the normals to the $(d-2)$-simplices $\sigma(0l)$ as deformation vectors, instead of edge vectors. The appearance of structure functions for the diffeomorphisms mirrors the appearance of structure functions for the commutator of hypersurface normals. Thus the commutator between diffeomorphism constraints might provide a toy model for systems with structure functions, which could be also tested for the spatial diffeomorphism constraints in 3D gravity.

On the other hand the structure function appearing for the commutator of two Hamiltonians is now somewhat simpler, namely just the inverse volume $1/V(0)$ of the `spatial' $(d-1)$-simplex.

Other choices for the diffeomorphism deformation vectors can be made and the algebra can be computed along the same lines. For instance we can choose, as in the Section \ref{4} for the 3D algebra, the edge vectors (from vertex $k$ to vertex $k'$) $(e_{kk'})^l=\delta^l_{k'}-\delta^l_k$ as deformation vectors. We will name the corresponding constraints $E(kk')$, which generate translations of the vertex $k$ in the direction $e_{kk'}$.

We find the following constraint algebra,
 \ba\label{sbda2}
 \{H(k), H(k') \} &=& \sum_l \tilde h(0)^{k l} E(k'l) -\sum_l \tilde h(0)^{k'l}E(kl) \nn\\
 \{E(kl),H(k')\}&=& (\delta^{k'}_{l} -\delta^{k'}_k) H(k) \nn\\
 \{E(kl),E(k'l')\}&=& (\delta^{k'}_l-\delta^{k'}_{k} )E(kl') -(\delta^k_{l'}-\delta^k_{k'}) E(k'l) -\delta^{k'}_{l} E(kk')+\delta^k_{l'}E(k'k)\, . \q\q
 \ea
Regarding the appearance of structure functions  this reflects now the continuum Poisson brackets. The last relation in (\ref{sbda2}) describes more generally the commutation relations between constraints generating translations of the vertices along the edge vectors (including `time like'  directions $(e_{0k})$). Constraints based on such deformation vectors would lead to an algebra with structure constants instead of structure functions.

So far we have been working with a set of constraints which is over--complete, even locally: at each vertex we have $d$ Hamiltonian constraints from the $d$ subsimplices of dimension $(d-1)$ meeting at this vertex. Furthermore we have $d\times (d-1)$ diffeomorphism constraints ($(d-1)$ diffeomorphism constraints per $(d-1)$--dimensional subsimplex) per vertex. As a vertex can be displaced only in $d$ directions, the set of Hamiltonians should be sufficient in the generic case. This will give $(d+1)\times d$ constraints which will still be a basis with global redundancy. If we subtract global rotations and global translations (or take into account that the length of an edge can be changed by translating both of its vertices), we will arrive at $\tfrac{1}{2} (d+1)\times d$ constraints, which is the number of edges. Thus we always have a totally constrained system.

We could just choose to work with the Hamiltonian constraints only, as is done in \cite{valentin1}. We so far restricted the Hamiltonian constraints to the normal $N(0)$, which we now relax. Thus we define the Hamiltonian constraints $H_c(kl)$ as deformations of the vertex $k$ with deformation vector $- c(kl) N(l)$. Here $c(kl)$ is a normalization factor. Choosing $c(kl)=1$ we obtain an Abelian algebra, as in this case the entries of the deformation (co--) vector $-N(l)_j=\tilde \delta^l_j$ do not depend on the geometry. Another choice for the normalization factors is $c(kl)= d \, V$. Thus the lengths of the deformation vector $-\check N(l)=-c(kl) N(l)$ is the `spatial volume' $V(l)$. This corresponds to a density two Hamiltonian, which is the choice of \cite{valentin1}. Naming the corresponding constraints $\tilde H(kl)$ the algebra is found to be
\ba\label{hamalg}
\{\tilde H(kl), \tilde H(k'l')\}&=& d\, V \left( \tilde g^{kl} \tilde H(k'l') - \tilde g^{k'l'} \tilde H(kl)\right) \q .
\ea

All the algebras we have discussed in this Section hold on the boundary of a simplex in any dimension, starting with the two--dimensional boundary of the tetrahedron\footnote{For the boundary of a triangle we would need to associate the configuration variables to vertices. This is only possible if we add fields beside the metric field.}. In this sense the algebras are universal. We now show how they are modified in the presence of the cosmological constant.

\section{Homogeneously curved simplices} \label{sec:curved}

Here we discuss gravity with a cosmological constant. We consider again the boundary of a $d$--simplex. For $(2+1)$D this can be generalized to arbitrary triangulations, while for $(3+1)$D we have again to restrict to the topological sector (e.g. triangulations describing stacked spheres, of which the boundary of a simplex is the simplest). Also we are going to discuss the theory in scalar variables, as so far a phase space description with (loop quantum gravity like) connection variables is not available.

Homogeneously curved simplices provide an improved discretization of gravity with a cosmological constant \cite{bahrdittrich2,newangle}. In particular in 3D, whereas the discretization provided by standard Regge calculus breaks diffeomorphism symmetry \cite{dittrichreview,bahrdittrich1}, the discretization with homogeneously curved simplices is symmetry preserving and triangulation independent \cite{bahrdittrich2,newangle}. The generalized Regge action associated to one homogeneously curved $d$--simplex is given as (we will restrict  to positive curvature, but all results can be easily generalized to negative curvature)
\ba\label{h1}
S_\sigma= \sum_{(ij)} V(ij)(\pi-\theta^\kappa(ij)) + (D-1)\kappa V \q .
\ea
Here we describe a space of constant section curvature: the sphere with radius $R=\frac{1}{\sqrt{\kappa}}$. The associated cosmological constant is $\Lambda=\frac{(d-1)(d-2)}{2}\kappa$. In (\ref{h1}) $\theta^\kappa(ij)$ is the dihedral angle at $\sigma(ij)$ in the curved geometry as a function of the lengths of the edges.  Also $V$ and $V(ij)$, which denote the volume of the simplex and of its $(d-2)$-subsimplices $\sigma(ij)$ respectively, have to be understood as functions of the lengths. The sum is over ordered pairs $(mn)$, $m<n$.

There is also a corresponding first order action \cite{newangle} to (\ref{h1}) given by
\ba\label{h2}
S^{fo}_\sigma= \sum_{(ij)} V(ij)(l)(\pi-\theta(ij)) + (D-1)\kappa V^\kappa(\theta)
\ea
where now the $\theta(ij)$ are independent variables and the volume $V^\kappa$ of $\sigma$ is a function of the dihedral angles. This shows that the dihedral angles $\theta(ij)$ and the volumina $V(ij)$ are conjugated variables. It can be also shown by varying  the second order action (\ref{h1})  with respect to the length variables and using the Schlaefli identity
\ba\label{h3}
(D-1)\kappa \delta V = \sum{(ij)} V(ij) \delta \theta^\kappa(ij) \q .
\ea
Hence we will adopt again the canonical Poisson brackets
\ba\label{h4}
\{V(ij),\theta(kl)\}=\delta_{(ij),(kl)} \q .
\ea
The constraints are similar to the flat case (see \cite{hoehn} how to derive them). Like in the case $\kappa=0$, they relate the extrinsic curvature to the intrinsic geometry, by fixing the angles $\theta_{ij}$ as the ones of the curved simplex determined by the lengths or the volumina $V(ij)$,
\ba\label{h5}
C_{(ij)}=\theta_{ij}-\theta_{ij}^\kappa \q .
\ea
The constraints are Abelian due to the Schlaefli identity (\ref{h3}). In the curved case we can also determine the volumina from the dihedral angles
\ba\label{h6bis}
\tilde C_{(ij)}= -V(ij)+V^\kappa(ij)
\ea
where now $V^\kappa(ij)$ are the volumina as functions of the dihedral angles. Those constraints are also Abelian. The geometric action of the constraints is that $C_{(ij)}$ shifts $V(ij)$, whereas $\tilde C_{(ij)}$ shifts the dihedral angle $\theta_{ij}$. This uniquely determines the flow of the constraints on the constraint hypersurface.

The constraints can be used to generate translations of the vertices in the homogeneously curved geometry, i.e. inside the sphere. We can describe those translations through deformation vectors $v$ tangential to the sphere at the vertex $k$ which chosen to be displaced. The associated constraint is given by
\ba\label{h7bis}
C_{k,v}=\sum_{(mn)} \delta_{k,v}(\theta(mn))\, \tilde C_{(mn)} \q ,
\ea
where the variation $\delta_{k,v}(\theta(mn))$ can be determined from the variations of the vertex vectors that we will introduce next. Thus for a given deformation we can find the constraints and choosing this deformation normal or tangential to the simplex boundary we obtain Hamiltonian or spatial diffeomorphism constraints.

We will not need the explicit form of these constraints to compute the constraint algebra. As before we will instead consider the commutators of the geometric deformations induced by translations of the vertices. Whereas for the flat simplex we used the affine metric to encode the geometry, we now use the vertex vectors for a spherical simplex. The simplex is embedded in a sphere of radius $\frac{1}{\sqrt{\kappa}}$, which itself is embedded into $\Rl^{d+1}$. Thus the vertices of this simplex can be described by $(d+1)$ vertex vectors $e_j$ pointing from the origin of $\Rl^{d+1}$ to the vertex $j=0,\dotsc,d$ on the sphere. We will denote by
\ba\label{h8bis}
G_{ij}&=&\cos(\sqrt{\kappa}l_{ij}) \;=\; \kappa\ \bigl(e_i\cdot e_j\bigr)
\ea
the length Gram matrix of the simplex. It gives access to the interior dihedral angles (and therefore the angle Gram matrix) as
\ba\label{h9}
\tilde G_{ij}\,=\, \cos \theta(ij) \,=\, -\frac{G^{ij}}{ \sqrt{G^{ii}G^{jj}}} \q .
\ea
Here $G^{ij}$ is the inverse of $G_{jk}$. The matrices $G, \tilde G$ allow to express the change of the dihedral angles under a variation of the vertex vectors $e_i$, see for instance \cite{bahrdittrich2,aristide3D}.

Translating a vertex $k$ by a vector $v$ just changes the corresponding vertex vector by
\ba\label{h10}
\delta_{k,v}(e_m)=\delta_{km} v \q .
\ea
For $v$ to describe an allowed deformation, it has to be tangential to the sphere at $e_k$, i.e. we have the condition $v\cdot e_k=0$.

Let us first choose deformation vectors that describe normal deformations to a given subsimplex. We fix the vertex $j=0$ and consider the normal to $\sigma(0)$. It is orthogonal to all vertex vectors $e_m$ with $m\neq 0$, ensuring that it is orthogonal to all edges adjacent to a given vertex $n\neq 0$ and not connecting to the vertex $0$. Such a normal is also tangential to the sphere at all vertices $n\neq 0$.

These considerations suggest the dual basis (see for instance \cite{aristide3D}) $(N(k))$ defined by
\ba\label{h11}
N(k)\cdot e_j =\delta_{kj} \q .
\ea
We can then take as the (outward pointing) normal $-\hat N(0)$ where the hat denotes normalization to one.

We will first consider the commutator of two Hamiltonians $H(k), H(k')$ at vertices $k, k'\neq 0$. This will suggest a choice for the vectors describing spatial diffeomorphisms. Reproducing the arguments of \eqref{n1},
\be
\left[ \delta_{H(k')} \circ \delta_{H(k)} - \delta_{H(k)} \circ \delta_{H(k')} \right] e_m = \delta_{km}\, \delta_{H(k')}(-\hat{N}(0)) - \delta_{k'm}\, \delta_{H(k)}(-\hat{N}(0)),
\ee
we see that we have to consider the change of the deformation vector $-\hat{N}(0)$ under the translation of the vertices $k$ and $k'$. For a general variation $\delta$ acting on the dual vector $\hat{N}(p)$ we have
\ba\label{h12}
\delta(\hat N(p)\cdot  \hat N(p))&=&0\;=\; 2 \hat{N}(p)\cdot \delta(\hat N(p)) \nn\\
\delta(\hat N(p) \cdot e_k) &=& 0\;=\; \hat{N}(p) \cdot \delta(e_k) + \delta(\hat N(p)) \cdot e_k \q .
\ea
Inserting the ansatz
\ba
\delta(\hat N(p))&=&\alpha \hat N(p)+ \sum_{k\neq p} \beta_k e_k
\ea
into (\ref{h12}) we find that
\ba
\alpha\,=\,0 \q,\q\q \sum_m \beta_m H_{mk}(p) + \delta(e_k)\cdot \hat N(p)\,=\,0  \q .
\ea
Here $H_{mk}(p)$ is the length Gram matrix for the subsimplex $\sigma(p)$, i.e. excluding the vertex $p$. Its inverse has matrix elements
\ba\label{h13}
H^{mk}(p)=G^{mk}-\frac{G^{mp}G^{pk}}{G^{pp}} \q .
\ea
Note the similarity to the inverse of the induced affine metric for the flat simplex \eqref{h^ll}. Also we can now allow $m,k=p$ as the inverse matrix elements are zero for these choices.

Therefore the change of the normal $\hat{N}(p)$ is
\ba\label{h14}
\delta (\hat N(p))=-\sum_{k,m} H^{km}(p) \left( \delta(e_k) \cdot \hat N(p) \right)\, e_m \q .
\ea
Thus the normal $-\hat N(0)$ changes under a Hamiltonian deformation $H(k)$ at the vertex $k$ by
\ba\label{h15}
\delta_{H(k)}(-\hat N(0))&=&-\sum_m H^{km}(0)\, e_m \q.
\ea
The commutator between the two Hamiltonian deformations becomes
\be
\begin{aligned}
\left[ \delta_{H(k')} \,\delta_{H(k)} - \delta_{H(k)} \,\delta_{H(k')} \right] e_m &= -\delta_{km} \sum_l H^{k'l}(0)\,e_l + \delta_{k'm} \sum_l H^{kl}(0)\,e_l \\
&= -\delta_{km} \sum_l H^{k'l}(0)\,\bigl(e_l - H_{lk}(0)\,e_k\bigr) + \delta_{k'm} \sum_l H^{kl}(0)\,\bigl(e_l-H_{lk'}(0)\,e_{k'}\bigr).
\end{aligned}
\ee
In the second line, we have introduced the vectors $e_{kl} = e_l -H_{lk}(0) e_k$ (no summation over $k$).  These vectors are in the plane spanned by $e_l$ and $e_k$ and are also orthogonal to $e_k$. Thus the vector $e_{kl}$ is tangent to the edge between the vertices $k$ and $l$ at the vertex $l$. Consequently, it is natural to use it as a deformation vector at the vertex $k$ for a diffeomorphism that we denote $E(kl)$. Its action is
\be
\delta_{E(kl)} e_m = \delta_{km}\,\bigl(e_l - H_{kl}(0)\,e_k\bigr).
\ee
This way, we find the commutator of two Hamiltonian constraints 
\ba\label{h16}
\{H(k),H(k')\}&=&\sum_{l}  H^{kl}(0)\,E(k'l) - \sum_{l}  H^{k'l}(0)\, E(kl)\q,
\ea
which in form agrees with the corresponding Poisson bracket (\ref{sbda2}) under the replacement of the affine inverse metric $\tilde h^{kl}(0)$ with the length Gram matrix $H^{kl}(0)$. Note however that as structure functions expressed in the volumina $V(ij)$, they will in generally differ. This means that the algebra for curved simplices is a deformation of the one for flat simplices.

Having chosen the deformation vectors corresponding to the spatial diffeomorphisms we can determine the Poisson brackets between Hamiltonian and diffeomorphism constraints and among the diffeomorphisms. We can actually generalize the vectors $e_{kl}$ to $d_{kl} = e_l- G_{lk} e_k$, where $l,k$ can take the value $0$. The change of the deformation vector $d_{im}$ under a deformation $\delta$ is
\ba\label{h17}
\delta(d_{im})&=& \delta(e_m)-G_{mi}\, \delta(e_i) - \kappa\ e_i\,\bigl(\delta(e_m)\cdot e_i + e_m\cdot \delta(e_i)\bigr) \q .
\ea
The resulting hypersurface deformation algebra is given by 
\ba\label{h18}
\{H(k),H(k')\}&=&\sum_{l}  H^{kl}(0)E(k'l) - \sum_{l}  H^{k'l}(0) E(kl) \nn\\
\{E(kl), H(k')\}&=& (\delta^{k'}_l -\delta^{k'}_k G_{kl})H(k)  \nn\\
 \{E(kl),E(k'l')\}&=& (\delta^{k'}_l-\delta^{k'}_{k}  G_{kl} )E(kl') -(\delta^k_{l'}-\delta^k_{k'} G_{k'l'}) E(k'l) -\delta^{k'}_{l} G_{l'l}E(kk')+\delta^k_{l'}G_{l'l}E(k'k)\, . \nn
\ea
Notice that in the flat limit $\kappa\rightarrow 0$, $G_{ij}\to 1$ and we recover the algebra for the flat simplex. However for generic $\kappa$, we see that with our choice of deformation vectors for the spatial diffeomorphisms we get structure functions for the Poisson brackets between Hamiltonians and diffeomorphisms and among diffeomorphisms. As $G_{kl}=H_{kl}(0)=\cos(\sqrt{\kappa} l_{kl})$ for $k,l\neq 0$ the structure functions only involve the `spatial geometry'.

Thus, the details of the discrete hypersurface deformation algebra does depend on the dynamics, here whether the cosmological constant is included or not. This differs from the continuum, where this is not the case.  We conjecture that the algebra will in general depend on the geometry that is prescribed as a solution of the equations of motions.

\section{Discussion} \label{sec:discussion}

The Dirac's hypersurface deformation algebra can also be defined for discrete hypersurfaces. We discussed an explicit realization for 3D gravity in terms of connection and triad variables. Realizations of this algebra can also be constructed in higher dimensional gravity, if one restricts to triangulations that necessarily describe flat, or in the presence of a cosmological constant, homogeneously curved, geometries.

We can draw various lessons from this work:
\begin{itemize}
\item One needs to be careful to which objects to assign the Hamiltonian and diffeomorphism constraints. Two choices which occurred so far in the literature are (a) to associate the constraints to the vertices in the discretization (i.e. faces in the dual for canonical 3D gravity or 3D cells in the dual for canonical 4D gravity), as is natural if the emphasis is on the action of the constraints as generators of vertex translations. The other option (b), used in loop quantum gravity \cite{qsd} is to associate the constraints to the top--dimensional simplices in the spatial discretization, i.e. polygonal cells in 3D gravity and 3D cells in 4D gravity (corresponding to vertices in the dual). This choice was critized in \cite{immirzi} from a discrete gravity perspective. Indeed in general the two options might constrain different numbers of degrees of freedom. A  counting of the constraints might however be misleading as redundancies might occur, a point missed in \cite{bander}.

  In this work we dealt with an overcomplete basis of constraints, associated to (vertex--dual vertex) pairs. Starting with this choice both options (a) and (b) can be discussed using some averaging prescription for the constraints. A precise understanding of the underlying geometry can help to select the better option and to avoid either over--constraining the theory or missing some constraints.

\item The details of the constraint algebra depend on the precise geometric definition of the constraints. For the Hamiltonian one has to describe the choice of a normal, for the diffeomorphism constraints tangential vectors to the hypersurface have to be selected. This choice influences in which relations structure functions will occur. 

It is possible to avoid structure functions (at least for the simplex boundary), by either using the Abelian deformations (\ref{linC}) and the one with only rescaled Hamiltonian constraints discussed above (\ref{hamalg}), 
 or deformations along the edge vectors (for the flat case). These choices do however not carry fully the interpretation of the continuum hypersurface deformation algebra.

\item The continuum constraints lead to a `linear representation' of Dirac's hypersurface deformation algebra. We have seen that in addition, also terms of higher order in the constraints might occur. These terms do not lead to any flow on the constraint hypersurface. But such terms might become relevant if gravity is coupled to other fields. In this case one might have to reduce to a flat sector, even for 3D gravity.

\item The Dirac algebra is universal in the continuum, e.g. does for instance not depend on whether a cosmological constant is present or not. We have seen that in the presence of the cosmological constant the discrete constraint algebra is deformed. This is due to the discretization: the structure functions refer to the geometry expressed via the lengths and volumina of finite size building blocks. The geometry of these building blocks differs in the flat and homogeneously curved case. Thus, should it be possible to define constraints for discrete 4D gravity, including inhomogeneously curved geometries, one has to expect an even more complicated algebra. (This might be possible if the constraints are non--local \cite{lincoarse,cylindrical}). However in the limit of very small discretization scale this dependence should disappear, as in this case the local geometry is almost flat.

\end{itemize}

Having clarified the classical representations for the hypersurface deformation algebra we prepared ground for investigating quantum representations. These seem to be in reach for the 3D theory: On the one hand one should investigate the algebra of quantum constraints in \cite{valentin1} and how this reflects the classical relations (\ref{hamalg}). One should be aware that terms of higher order in the constraints could appear. The algebra has structure functions, thus there is the questions of how these are ordered with respect to the constraints on the right hand side of (\ref{hamalg}).

On the other hand one could consider quantization of the diffeomorphism constraints and the density one Hamiltonian constraints, following section \ref{9} and adapting the techniques of \cite{qsd,qsd3D} to a simplicial context. Here we clarified that the Thiemann trick is also available as an exact identity in an simplicial phase space. Indeed, in this context one has to be careful to adjust numerical coefficients, that might differ from the continuum formulas. Similar considerations in 4D are also possible.

Considering density one Hamiltonians, density two Hamiltonians or the flatness constraints will influence how degenerate configurations are treated. This is still an open question of research \cite{witten}.  This question is related to the choice of boundary conditions for `zero spatial slices' and also the question whether one has necessarily to sum over orientations \cite{carloorient}.
As pointed out in \cite{valentin1} the boundary conditions for the density two Hamiltonian indeed impose that the physical wave functions includes both orientations. Thus investigating other choices for the density could uncover general mechanisms, that might also be applicable to bouncing cosmologies or singularity resolutions \cite{bojowald}.

Furthermore for the quantization of the density one Hamiltonians factor ordering ambiguities arise that are analogous to those in the 4D theory. These might be much easier to resolve in the 3D theory, so that lessons can be drawn for the 4D case.

As we have discussed in 4D we can consider a topological sector of the discretized theory that admits exact constraints. These allow to define a hypersurface deformation algebra, which we here considered for the boundary of a simplex. Let us point out that even finding the physical wave function for the simplex would mean considerable progress for the 4D theory: As in the 3D theory this physical wave function would constitute the amplitude for the simplex form which we can built the path integral, that is spin foam model, of simplicial gravity. Such a physical wave function would also include information on the measure factors, which in 4D are not specified even for configurations describing the classically flat sector \cite{steinhausM}. This of course would strengthen very much the connection between canonical loop quantum gravity and spin foam models, and clarify the role of diffeomorphism symmetry and the related Slavnov--Taylor identities along the lines of \cite{recursion}.

To this end it would be beneficial to investigate more in detail the simplicial version of the Hamiltonian constraints \cite{qsd}, which has been started in \cite{valentin4D}. The question arises whether one can find a unique quantum representation of a given simplex boundary deformation algebra or whether there exist more than one representation. An answer to this question would be useful for canonical quantization as well as the path integral approach. 

\section*{\small Acknowledgements}
We thank Laurent Freidel for discussions which motivated this work. Research at Perimeter Institute is supported by the Government of Canada through Industry Canada and by the Province of Ontario through the Ministry of Research and Innovation.

\appendix
\section{Evaluation of Poisson brackets} \label{app:calculations}

Here we provide some details on the generic calculation of the Poisson bracket \eqref{pb1}. The edges associated to the pair $(f,v)$ are labeled with Greek letters $\alpha, \beta=1,2$, and those associated to $(f',v')$ are labeled with $\alpha', \beta'=3,4$.

\medskip

{\it (a)\ }
Let us consider the case that $A=E_\alpha$. Then the first term in (\ref{pb1}) is given by
\be\label{pb4}
h^l_{fv}    \,\{ E_\alpha^l, h^m_{f'v'}\}  \,B^m =
o_{\alpha f'}\,   h^0_{f'}   h^k_{fv}  (\Ad_{f':v'v}(B) )^k \, +  \frac{1}{2} o_{\alpha f'}\, h^p_{f'v}  \, h^k_{fv}  \, \epsilon^{pkn} \,  (\Ad_{f':v'v}(B) )^n \q .
\ee
We need to split $\tilde B = (\text{Ad}_{f':v'v}(B) )$ into tangential and normal components at $(f,v)$,
\ba
\tilde B  &=& ( \tilde B \cdot E_\beta) Q_{fv}^{\beta\gamma}  E_{\gamma} + (\tilde B \cdot n_{fv}) n_{fv}\;.
\ea
Thus
\be
h^k_{fv}  (\Ad_{f':v'v}(B) )^k =  ( \tilde B \cdot E_\beta) Q_{fv}^{\beta\gamma} D_{\gamma}^{fv} +  (\tilde B \cdot n_{fv}) H^{fv} \q .
\ee

Whereas in the first term of the right hand side of \eqref{pb4} we have $h_{f'}^0$ appearing which is equal to $1$ on the constraint hypersurface, the second contribution of (\ref{pb4}) a term quadratic in the flatness constraints. (Note that this term does not generate a flow on the constraint hypersurface.) To split these into Hamiltonian and diffeomorphisms we could write
\begin{alignat}{3}\label{hdecomp}
 h^k_{fv}   &= ( h_{fv} \cdot E_\beta)\ Q^{\beta\gamma}_{fv}\ E_{\gamma}^k  &+&\ ( h_{fv} \cdot n_{fv}) \ n^k_{fv} &=&\,  D^{fv}_{\beta} Q^{\beta\gamma}_{fv} E_{\gamma}^k +H^{fv} \, n^k_{fv} \nn\\
 h^p_{f'v} &= ( h_{f'v} \cdot E_{\rho'}) Q^{\rho'\sigma'}_{f'v} E_{\sigma'}^p  &+&\, ( h_{f'v} \cdot n_{f'v}) \, n^p_{f'v}\  &=&\,  D^{f'v}_{\rho'} Q^{\rho'\sigma'}_{f'v} E_{\sigma'}^p +H^{f'v} \, n^p_{f'v}  \q .
\end{alignat}

This would however introduce a third set of constraints at $(f'v)$, whose splitting is not adjusted to $B$ and (in the general case) not to $n_{fv}$. 
We can  transport the constraints to $v'$ using the following rewriting
\ba
\epsilon^{pkn} h^p_{f'v} (\Ad_{f':v'v}(B))^n
&=&
(-2)\, \tr( h_{f':v'v} h_{f':vv'} \, [T^k \, , h_{f':v'v}B h^{-1}_{f':v'v} ] )   \nn\\
&=&
(-2)\tr( [B,h_{f'v'}]\, (h^{-1}){f'v'v} T^k h_{f'v'v})) \nn\\
&=&
\epsilon^{npm} B^nh^p_{f'v'} ( \Ad_{(f')^{-1}:vv'}(T^k))^m \nn\\
&=&
\epsilon^{npm} B^n \, (D^{f'v'}_{\beta'} Q^{\beta'\gamma'}_{f'v'} E^p_{\gamma'} + H^{f'v'} n^p_{f'v'} )\,( \Ad_{(f')^{-1}:vv'}(T^k))^m \, .\q\q\q
\ea

Collecting all the terms, the first term in (\ref{pb1}) is given by (for $A=E_\alpha$)
\begin{multline}\label{pb4z}
h^l_{fv}  \,\{ E_\alpha^l, h^m_{f'v'}\}  \,B^m =
o_{\alpha f'}\,   h^0_{f'}   \,\left( ( \tilde B \cdot E_{\beta} \,\, Q_{fv}^{\beta\gamma} ) D_{\gamma}^{fv} +  (\tilde B \cdot n_{fv}) H^{fv} \right) \\
 + \tfrac{1}{2} o_{\alpha f'}\,
 \epsilon^{npm} B^n \, \left(D^{f'v'}_{\beta'} Q^{\beta'\gamma'}_{f'v'} E^p_{\gamma'} + H^{f'v'} n^p_{f'v'} \right)\\
\times \left(   D^{fv}_{\beta} Q^{\beta\gamma}_{fv}     ( \Ad_{(f')^{-1}:vv'}  E_{\gamma})^m +H^{fv} \,(\Ad_{(f')^{-1}:vv'} n_{fv} )^m \right)  \, . \q\q\q
\end{multline}

For $B$ equal to the normal $n_{f'v'}$  we just need to consider the term
\ba
 \epsilon^{npm} n_{f'v'}^n E^p_{\gamma'}
 &=& \frac{1}{|E_3\times E_4|}\left( (E_3 \cdot E_{\gamma'} )\, E_4^m - (E_4 \cdot E_{\gamma'}) \, E_3^m\right)  \nn\\
 &=&  |E_3\times E_4| \,\, \epsilon_{\gamma' \delta'} Q^{\delta'\alpha'}_{f'v'} E_{\alpha'}^m \q
\ea
with $\epsilon_{\gamma'\delta'}$ totally antisymmetric and $\epsilon_{34}=-\epsilon_{43}=1$. Using furthermore
\ba
\epsilon_{\gamma'\delta'} Q^{\beta'\gamma'}_{f'v'} Q^{\delta'\alpha'}_{f'v'} &=& \det(Q^{-1}_{f'v'}) \epsilon^{\beta'\alpha'} \,=\, \frac{1}{|E_3\times E_4|^2}  \epsilon^{\beta'\alpha'}
\ea
(with $\epsilon^{34}=-\epsilon^{43}=1$) we obtain for the case $A=E_\alpha$ and $B=n_{f'v'}$
\begin{multline}
h^l_{fv}    \,\{ E_\alpha^l, h^m_{f'v'}\}  \,n_{f'v'}^m =
o_{\alpha f'}\,   h^0_{f'}   \,\left( ( \tilde n_{f'v'} \cdot E_{\beta} \,\, Q_{fv}^{\beta\gamma} ) D_{\gamma}^{fv} +  (\tilde n_{f'v'} \cdot n_{fv}) H^{fv} \right) \\
+ \tfrac{1}{2} o_{\alpha f'}\,
  \frac{1}{  \sqrt{\det Q_{f'v'}}}  \epsilon^{\beta'\alpha'}  D^{f'v'}_{\beta'}  E_{\alpha'}^m
  \left(   D^{fv}_{\beta} Q^{\beta\gamma}_{fv}     ( \Ad_{(f')^{-1}:vv'}  E_{\gamma})^m +H^{fv} \,(\Ad_{(f')^{-1}:vv'} n_{fv} )^m \right)  \;.
\end{multline}
Similarly we find for $A=E_\alpha$ and $B=E_{\alpha'}$
\ba
h^l_{fv} \,\{ E_\alpha^l, h^m_{f'v'}\}  \,E_{\alpha'}^m \!\!\!\!
&=&
o_{\alpha f'}\,   h^0_{f'}   \,\left( ( \tilde E_{\alpha'} \cdot E_{\beta} \,\, Q_{fv}^{\beta\gamma} ) D_{\gamma}^{fv} +  (\tilde E_{\alpha'} \cdot n_{fv}) H^{fv} \right) \, +\nn\\&&
 \tfrac{1}{2} o_{\alpha f'}\,
 \sqrt{\det Q_{f'v'}} \, Q^{\beta'\delta'}_{f'v'} \epsilon_{\alpha'\delta'} \left( D^{f'v'}_{\beta'} \, n^m_{f'v'} \,-\, H^{f'v'} E^m_{\beta'}\right) \nn\\&&
\times \left( D^{fv}_{\beta} Q^{\beta\gamma}_{fv}     ( \Ad_{(f')^{-1}:vv'}  E_{\gamma})^m +H^{fv} \,(\Ad_{(f')^{-1}:vv'} n_{fv} )^m \right) \, . \q\q\q
\ea

\medskip

{\it (b)\ } The other case is that $A=n_{fv}$, for which we can write
\be\label{pb5}
h^l_{fv}   \frac{\partial n^l_{fv}}{\partial E_\alpha^k} \,\{ E_\alpha^k, h^m_{f'v'}\}  \,B^m \!\!
= -o_{\alpha f'} h^0_{f'}  \,  Q_{fv}^{\alpha\beta} \, D^{fv}_\beta \, (n_{fv}  \cdot\tilde B)
-\tfrac{1}{2} o_{\alpha f'} \,  Q_{fv}^{\alpha\beta} D^{fv}_\beta \,  \epsilon^{pkn}  h^p_{f'v}      n^k_{fv} \, \tilde B^n \; .\q\q
\ee
We have to rewrite the second term of the right hand side above using
\ba
 \epsilon^{pkn}  h^p_{f'v}      n^k_{fv} \, \tilde B^n  &=&
 \epsilon^{npm} B^n  \, (D^{f'v'}_{\gamma'} Q^{\gamma'\delta'}_{f'v'} E^p_{\delta'} + H^{f'v'} n^p_{f'v'} )\, ( \text{Ad}_{(f')^{-1}:vv'}(n_{fv}))^m \, .\q\q
\ea
We find for $A=n_{fv},\,\,B=n_{f'v'}$
\begin{multline}
h^l_{fv}   \frac{\partial n^l_{fv}}{\partial E_\alpha^k} \,\{ E_\alpha^k, h^m_{f'v'}\}  \,n_{f'v'}^m = -o_{\alpha f'} h^0_{f'}  \,  Q_{fv}^{\alpha\beta} \, D^{fv}_\beta \, (n_{fv}  \cdot\tilde n_{f'v'})\\
-\tfrac{1}{2} o_{\alpha f'} \,\frac{1}{  \sqrt{\det Q_{f'v'}}}  \epsilon^{\beta'\gamma'}  D^{f'v'}_{\beta'}
  E_{\gamma'}^m\ D^{fv}_\beta \,  Q_{fv}^{\beta\alpha}  \, ( \Ad_{(f')^{-1}:vv'}(n_{fv}))^m
 \;,
\end{multline}
and for $A=n_{fv},\,\,B=E_{\alpha'}$
\begin{multline}
h^l_{fv}   \frac{\partial n^l_{fv}}{\partial E_\alpha^k} \,\{ E_\alpha^k, h^m_{f'v'}\}  \,E_{\alpha'}^m = -o_{\alpha f'} h^0_{f'}  \,  Q_{fv}^{\alpha\beta} \, D^{fv}_\beta \, (n_{fv}  \cdot\tilde E_{\alpha'}) \\
-\tfrac{1}{2} o_{\alpha f'} \,
\sqrt{\det Q_{f'v'}} \, Q^{\gamma'\delta'}_{f'v'} \epsilon_{\alpha'\delta'} \left( D^{f'v'}_{\gamma'} \, n^m_{f'v'} \,-\, H^{f'v'} E^m_{\gamma'}\right) D^{fv}_\beta \,  Q_{fv}^{\beta\alpha}  \, ( \Ad_{(f')^{-1}:vv'}(n_{fv}))^m
 \;.
\end{multline}

\medskip

{\it (c)\ } For the second term in (\ref{pb1}) the same discussion applies. This will lead to terms linear in the constraints at $(f'v')$ and to terms quadratic  in constraints at $(f'v')$ and $(fv)$. Cancellations might occur between the first two terms in  (\ref{pb1}) (for the part quadratic in the constraints if we are not considering $f=f'$ and $v=v'$).

\medskip

{\it (d)\ }The last term in  (\ref{pb1})  is only non--vanishing for $v=v'$.
\ba
\delta_{vv'} h^l_{fv} \, h^m_{f'v'}   \frac{\partial A^l}{\partial E_\alpha^k}   \frac{\partial B^m}{\partial E_{\alpha'}^p} \{ E_\alpha^k\, ,\, E_{\alpha'}^p\}
&=& \sum_{\alpha,\gamma} \delta_{vv'} h^l_{fv} \, h^m_{f'v}   \frac{\partial A^l}{\partial E_\alpha^k}   \frac{\partial B^m}{\partial E_{\alpha'}^p} \delta_{\alpha\alpha'} \epsilon^{kpn}E^n_\alpha \q\q
\ea
where $\delta_{\alpha\gamma}=1$ if $e_\alpha=e_\gamma$ and vanishing otherwise. This will give only terms quadratic in the constraints, thus cancellations might occur with the quadratic constraint terms originating from the first two terms in (\ref{pb1}). Again we can decompose the face holonomies and apply a similar strategy as in {\it (a)}.
\begin{enumerate}
\item $A= E_\alpha, \,\,B=E_{\alpha'}$:
\be
\begin{aligned}
h^l_{fv} \, h^m_{f'v} \{ E_\alpha^l\, ,\, E_{\alpha'}^m\} &= \delta_{\alpha \alpha'}   h^l_{fv} \, h^m_{f'v}    \epsilon^{lmn}E^n_\alpha \\
&= \delta_{\alpha \alpha}   \left( D^{f'v}_{\gamma'} Q^{\gamma'\delta'}_{f'v} E^m_{\delta'} \,+\, H^{f'v} n^m_{f'v} \right) \sqrt{\det Q_{fv}}\, Q_{fv}^{\gamma'\beta'} \epsilon_{\alpha\beta'} \left(  D^{fv}_{\gamma'}  n^m_{fv} - H^{fv} E^m_{\gamma'} \right).
\end{aligned}
\ee

\item $A= n_{fv}, \,\, B=E_{\alpha'}$:
\be
h^l_{fv} \, h^m_{f'v}\,\frac{\partial n^l_{fv}}{\partial E_\alpha^k}   \{ E_\alpha^k\, ,\, E_{\alpha'}^m\}
=
-\delta_{\alpha \alpha'}
\left( D^{f'v}_{\gamma'} Q^{\gamma'\delta'}_{f'v} E^m_{\delta'} \,+\, H^{f'v} n^m_{f'v} \right) \frac{ \epsilon^{\gamma \delta}}{  \sqrt{\det Q_{fv}}  } (Q_{fv})_{\delta\alpha} Q^{\alpha\beta}_{fv}\, D^{fv}_{\beta}E_{\gamma}^m.
\ee

\item \label{case3} $A= n_{fv}, \,\, B=n_{f'v}$:
\be
h^l_{fv} \, h^m_{f'v}   \frac{\partial n^l_{fv}}{\partial E_\alpha^k}   \{ E_\alpha^k\, ,\, E_{\alpha'}^p\}  \frac{\partial n^m_{f'v}}{\partial E_{\alpha'}^p}
=\delta_{\alpha \alpha'}  \,  D^{fv}_\beta \, D^{f'v}_{\beta'} \,
(Q_{fv})_{\gamma\alpha}Q^{\alpha\beta}_{fv}  Q^{\alpha'\beta'}_{f'v}
\frac{1} {\sqrt{\det Q_{fv}} }  \, \epsilon^{\delta\gamma} \,\, E_{\delta}  \cdot     n_{f'v} \; .
\ee
\end{enumerate}

If we consider a three--valent vertex with outgoing edges the Gau\ss~constraint imposes $E_1+E_2+E_3=0$ for the three edges at $v$. Hence the normals to the three faces coincide and there will occur various simplifications, for instance case \ref{case3} above will lead to a vanishing result.


In summary the Poisson brackets between constraints based at $(fv)$ and $(f'v')$ can be expressed as combinations of constraints at $(fv)$ and $(f'v')$ again. In general one has to expect also terms quadratic in the constraints. These terms do not lead to a flow on the constraint hypersurface, thus there is no geometric interpretation for these quadratic terms. The terms linear in the constraints are however universal and can be indeed derived from geometrical considerations as we will comment on in sections \ref{5} and \ref{simplex}.

\section{Affine coordinates and metric for a simplex}
\label{A}

It is convenient to use the affine (or barycentric) coordinates introduced in \cite{sorkin} for Regge calculus
(see also \cite{dsl,areaangle} which use affine coordinates to express length derivatives of the dihedral angles and to proof various geometric identities) . Affine coordinates for a $d$--simplex are defined as follows.
Let $\vec{v}_j,\, j=0,\ldots,d$ be vectors from some arbitrarily chosen point in $\mathbb{R}^d$ to the $d+1$ vertices of an $d$--simplex $\sigma$.
Then
\ba
{\mathbf e}_j \equiv \vec{v}_j- \frac{1}{d+1}\sum_{k=0}^d \vec{v}_k
\ea
define an overcomplete affine basis. Hence the affine coordinates $\tilde x^j$ of a vector $\vec{x}=\sum_j \tilde x^j {\mathbf e}_j$
are not unique. However uniqueness can be obtained by imposing the additional condition $\sum_{j}\tilde x^j=0$.

A dual affine basis ${\mathbf e}^j$ is defined by
\ba
\mathbf e_j \cdot \mathbf e^k = \tilde \delta^k_j \equiv \delta^k_j-\frac{1}{d+1}
\ea
where $\delta^k_j$ is the Kronecker delta. The dual basis satisfies $\sum_j {\mathbf e}^j=0$.


The key quantity for our calculations is the metric tensor in these coordinates. It can be shown (by contraction with the edge vectors) that
the affine components of the metric tensor are \cite{sorkin}
\ba\label{affinemetric}
\tilde g_{ij}=-\tfrac{1}{2} \sum_{k, l} l^2_{kl} \, \tilde \delta^k_i \, \tilde \delta^l_j.
\ea

Laplace's formula for the determinant of $\tilde g_{ij}$ gives the squared of the $d$-volume of the simplex,
\ba\label{volume1}
V^2=\frac{1}{(d!)^{3}}\, \tilde \epsilon^{k_0 k_1\cdots k_{d-1}}\,\tilde\epsilon^{l_0 l_1 \cdots l_{d-1}}\,\tilde g_{k_0 l_0}\cdots \tilde g_{k_{d-1} l_{d-1}},
\ea
where the affine epsilon tensor is given by
\ba
\tilde \epsilon^{j_0\cdots j_{d-1}} =
\begin{cases}
+1\,\,& \mbox{if the permutation}\,\, j_0 \cdots j_{d-1} j_{d} \,\, \mbox{is even} \\
-1\,\,  &\mbox{if the permutation}\,\, j_0 \cdots j_{d-1} j_{d} \,\, \mbox{is odd}
\end{cases}
\ea
and is vanishing if $\{j_0 \cdots j_{d-1}\}$ can not be completed to a permutation of $\{0,\ldots,d\}$.
The inverse affine metric components $\tilde g^{ij}$ are defined by
$\tilde g_{kl} \, \tilde g^{li}=\tilde g^{il} \, \tilde g_{lk}=\tilde \delta^i_k$, and one can use (\ref{volume1})
to write
\ba\label{inverse metric}
\tilde g^{ij}
\,=\,\frac{d}{(d!)^3}\,\frac{1}{V^2}\,\tilde\eps^{i k_1\cdots k_{d-1}}\,\tilde\eps^{j l_1 \cdots l_{d-1}}\,\tilde g_{k_1 l_1}\cdots \tilde g_{k_{d-1} l_{d-1}} \,=\,-\frac{1}{V^2} \frac{\partial V^2}{\partial l^2_{ij}}
\ea
from which
\ba
\delta V = \frac{1}{2} V \tilde g^{kl} \, \delta(\tilde g_{kl})
\ea
for the variation of the volume follows.

The diagonal components of the inverse metric $\tilde g^{ii}$ are
proportional to the $(n-1)$-volumes of the $(i)$-subsimplices,
\ba
\tilde g^{ii}=\frac{1}{d^2}\frac{V(i)^2}{V^2}.
\ea

The edge vectors $(e_{mn})^l$ (from the vertex $m$ to the vertex $n$) are given as $(e_{mn})^l=\delta^l_n-\delta^l_m$. The (inward pointing) normals $N(i)$  to a subsimplex $\sigma(i)$ are given as
$N(i)_k = \tilde \delta^i_k$, and their norm is $|N(i)|^2=\tilde g^{ii}$. Thus one finds as a formula for the interior dihedral angles
\ba
\cos\theta(ij)= -\frac{ N(i)\cdot N(j) } {|N(i)|\, |N(j)|}  \,=\, -\frac{\tilde g^{ij}}{\sqrt{\tilde g^{ii} \, \tilde g^{jj}}} \q .
\ea



{\small
\begin{thebibliography}{99}
\parskip -2pt

\bibitem{teitelboim}
C.~Teitelboim, ``How commutators of constraints reflect the spacetime structure,'' Annals of Physics {\bf 79} (1973) 542

\bibitem{dirac}
P.~A.~M.~Dirac, ``Lectures on Quantum Mechanics''  (Belfer Graduate School of Science, Yeshiva University,  New York 1964)

\bibitem{habitat}
T.~Thiemann,
  ``QSD III: Quantum constraint algebra and physical scalar product in  quantum
  general relativity,''
  Class.\ Quant.\ Grav.\  {\bf 15} (1998) 1207
  [arXiv:gr-qc/9705017].
  J.~Lewandowski and D.~Marolf,
  ``Loop constraints: A habitat and their algebra,''
  Int.\ J.\ Mod.\ Phys.\  D {\bf 7} (1998) 299
  [arXiv:gr-qc/9710016].
 R.~Gambini, J.~Lewandowski, D.~Marolf and J.~Pullin,
  ``On the consistency of the constraint algebra in spin network quantum
  gravity,''
  Int.\ J.\ Mod.\ Phys.\  D {\bf 7} (1998) 97
  [arXiv:gr-qc/9710018].


\bibitem{nicolai}
  H.~Nicolai, K.~Peeters and M.~Zamaklar,
  ``Loop quantum gravity: An Outside view,''
  Class.\ Quant.\ Grav.\  {\bf 22} (2005) R193
  [hep-th/0501114].

\bibitem{vara}
 C.~Tomlin and M.~Varadarajan,
  ``Towards an Anomaly-Free Quantum Dynamics for a Weak Coupling Limit of Euclidean Gravity,''
  arXiv:1210.6869 [gr-qc].


\bibitem{qsd}
T.~Thiemann,
  ``Anomaly-free formulation of non-perturbative, four-dimensional  Lorentzian
  quantum gravity,''
  Phys.\ Lett.\  B {\bf 380} (1996) 257
  [arXiv:gr-qc/9606088].
 ``Quantum spin dynamics (QSD),''
  Class.\ Quant.\ Grav.\  {\bf 15} (1998) 839
  [arXiv:gr-qc/9606089].


\bibitem{loll}
 R.~Loll,
  ``On the diffeomorphism-commutators of lattice quantum gravity,''
  Class.\ Quant.\ Grav.\  {\bf 15} (1998) 799
  [arXiv:gr-qc/9708025].


\bibitem{dittrichreview}
 B.~Dittrich,
  ``Diffeomorphism symmetry in quantum gravity models,''  Adv.\ Sci.\ Lett.\ {\bf 2} (2009) 151,
 [arXiv:0810.3594 [gr-qc].]
%


\bibitem{bahrdittrich1}
 B.~Bahr and B.~Dittrich,
  ``(Broken) Gauge Symmetries and Constraints in Regge Calculus,''
  Class.\ Quant.\ Grav.\  {\bf 26} (2009) 225011
  [arXiv:0905.1670 [gr-qc]].


\bibitem{steinhaus1}
  B.~Bahr, B.~Dittrich and S.~Steinhaus,
  ``Perfect discretization of reparametrization invariant path integrals,''
  Phys.\ Rev.\ D {\bf 83} (2011) 105026
  [arXiv:1101.4775 [gr-qc]].



\bibitem{dittrichzako}
 B.~Dittrich,
  ``How to construct diffeomorphism symmetry on the lattice,''
  PoS QGQGS {\bf 2011} (2011) 012
  [arXiv:1201.3840 [gr-qc]].


\bibitem{hoehn}
 B.~Dittrich and P.~A.~H\"ohn,
  ``From covariant to canonical formulations of discrete gravity,''
  Class.\ Quant.\ Grav.\  {\bf 27} (2010) 155001
  [arXiv:0912.1817 [gr-qc]].
B.~Dittrich and P.~A.~H\"ohn,
  ``Canonical simplicial gravity,''
  Class.\ Quant.\ Grav.\  {\bf 29} (2012) 115009
  [arXiv:1108.1974 [gr-qc]].
 B.~Dittrich, P.~A.~Hoehn and ,
  ``Constraint analysis for variational discrete systems,''
  arXiv:1303.4294 [math-ph].


\bibitem{consistent}
R.~Gambini and J.~Pullin,
  ``Canonical quantization of general relativity in discrete space-times,''
  Phys.\ Rev.\ Lett.\  {\bf 90} (2003) 021301
  [arXiv:gr-qc/0206055].
 C.~Di Bartolo, R.~Gambini, R.~Porto and J.~Pullin,
  ``Dirac-like approach for consistent discretizations of classical
  constrained theories,''
  J.\ Math.\ Phys.\  {\bf 46} (2005) 012901
  [arXiv:gr-qc/0405131].

\bibitem{regge}
 T.~Regge,
  ``General relativity without coordinates,''
  Nuovo Cim.\  {\bf 19} (1961) 558.\\
  T.~Regge and R.~M.~Williams,
  ``Discrete structures in gravity,''
  J.\ Math.\ Phys.\  {\bf 41} (2000) 3964
  [arXiv:gr-qc/0012035].
%

\bibitem{bahrdittrich2}
 B.~Bahr and B.~Dittrich,
  ``Improved and Perfect Actions in Discrete Gravity,''
  Phys.\ Rev.\  D {\bf 80} (2009) 124030
  [arXiv:0907.4323 [gr-qc]].
%
\bibitem{witten3d} 
  E.~Witten,
  ``(2+1)-Dimensional Gravity as an Exactly Soluble System,''
  Nucl.\ Phys.\ B {\bf 311}, 46 (1988).


\bibitem{qsd3D}
 T.~Thiemann,
  ``QSD 4: (2+1) Euclidean quantum gravity as a model to test (3+1) Lorentzian quantum gravity,''
  Class.\ Quant.\ Grav.\  {\bf 15} (1998) 1249
  [gr-qc/9705018].


\bibitem{barrettcraneH}
J.~W.~Barrett and L.~Crane,
  ``An Algebraic interpretation of the Wheeler-DeWitt equation,''
  Class.\ Quant.\ Grav.\  {\bf 14} (1997) 2113
  [gr-qc/9609030].

\bibitem{valentin1}
  V.~Bonzom and L.~Freidel,
  ``The Hamiltonian constraint in 3d Riemannian loop quantum gravity,''
  Class.\ Quant.\ Grav.\  {\bf 28} (2011) 195006
  [arXiv:1101.3524 [gr-qc]].
%
 V.~Bonzom and E.~R.~Livine,
  ``A New Hamiltonian for the Topological BF phase with spinor networks,''
  J.\ Math.\ Phys.\  {\bf 53} (2012) 072201
  [arXiv:1110.3272 [gr-qc]].
   V.~Bonzom and A.~Laddha,
  ``Lessons from toy-models for the dynamics of loop quantum gravity,''
  SIGMA {\bf 8} (2012) 009
  [arXiv:1110.2157 [gr-qc]].

\bibitem{giddings}
S.~Giddings, J.~Abbott and K.~Kuchar,
  ``Einstein's theory in a three-dimensional space-time,''
  Gen.\ Rel.\ Grav.\  {\bf 16} (1984) 751.
\bibitem{thooft}
 G.~'t Hooft,
  ``Causality in (2+1)-dimensional gravity,''
  Class.\ Quant.\ Grav.\  {\bf 9} (1992) 1335.
\bibitem{witten}
 E.~Witten,
  ``Three-Dimensional Gravity Revisited,''
  arXiv:0706.3359 [hep-th].
\bibitem{carlip}
 S.~Carlip,
  ``Quantum gravity in 2+1 dimensions: The Case of a closed universe,''
  Living Rev.\ Rel.\  {\bf 8} (2005) 1
  [gr-qc/0409039].
\bibitem{meusburger}
 C.~Meusburger,
  ``Geometrical (2+1)-gravity and the Chern-Simons formulation: Grafting, Dehn twists, Wilson loop observables and the cosmological constant,''
  Commun.\ Math.\ Phys.\  {\bf 273} (2007) 705
  [gr-qc/0607121].
C.~Meusburger,
  ``Cosmological measurements, time and observables in (2+1)-dimensional gravity,''
  Class.\ Quant.\ Grav.\  {\bf 26} (2009) 055006
  [arXiv:0811.4155 [gr-qc]].



\bibitem{notes}
B.~Dittrich, PSI lectures on quantum gravity and quantum geometry, to appear

\bibitem{waelbroeck}
H.~Waelbroeck and J.~A.~Zapata,
  ``Translation Symmetry In 2+1 Regge Calculus,''
  Class.\ Quant.\ Grav.\  {\bf 10} (1993) 1923.
  ``$2+1$ Covariant Lattice Theory and t'Hooft's Formulation,''
  Class.\ Quant.\ Grav.\  {\bf 13} (1996) 1761
  [arXiv:gr-qc/9601011].

\bibitem{alexkarim}
K.~Noui and A.~Perez,
  ``Three dimensional loop quantum gravity: Physical scalar product and  spin
  foam models,''
  Class.\ Quant.\ Grav.\  {\bf 22} (2005) 1739
  [arXiv:gr-qc/0402110].




\bibitem{PR1}
 L.~Freidel and D.~Louapre,
  ``Ponzano-Regge model revisited. I: Gauge fixing, observables and
  interacting spinning particles,''
  Class.\ Quant.\ Grav.\  {\bf 21} (2004) 5685
  [arXiv:hep-th/0401076].





\bibitem{bander}
 M.~Bander,
  ``Hamiltonian lattice gravity. 1. Deformations of discrete manifolds,''
  Phys.\ Rev.\  D {\bf 36} (1987) 2297.


\bibitem{utrecht}
 J.~Eldering,
  ``The Polygon model for 2+1D gravity: The Constraint algebra and problems of quantization,''
  gr-qc/0606132.


\bibitem{valentin4D}
 V.~Bonzom,
  ``Spin foam models and the Wheeler-DeWitt equation for the quantum 4-simplex,''
  Phys.\ Rev.\ D {\bf 84} (2011) 024009
  [arXiv:1101.1615 [gr-qc]].


\bibitem{louapre}
L.~Freidel and D.~Louapre,
  ``Diffeomorphisms and spin foam models,''
  Nucl.\ Phys.\  B {\bf 662} (2003) 279
  [arXiv:gr-qc/0212001].




\bibitem{ashtekarbook}
A. Ashtekar, ``New Perspepectives in Canonical Gravity'' (Monographs and Textbooks in Physical Science, Bibliopolis, Napoli, 1988)



\bibitem{qsd7}
 T.~Thiemann,
  ``Quantum spin dynamics (QSD): 7. Symplectic structures and continuum lattice formulations of gauge field theories,''
  Class.\ Quant.\ Grav.\  {\bf 18} (2001) 3293
  [hep-th/0005232].

\bibitem{immirzi}
 G.~Immirzi,
  ``Quantum gravity and Regge calculus,''
  Nucl.\ Phys.\ Proc.\ Suppl.\  {\bf 57} (1997) 65
  [arXiv:gr-qc/9701052].

\bibitem{teitelboim2}
S.~A.~Hojman, K.~Kuchar and C.~Teitelboim,
  ``Geometrodynamics Regained,''
  Ann.\ Phys.\  {\bf 96} (1976) 88.


\bibitem{areaangle}
B.~Dittrich and S.~Speziale,
  ``Area-angle variables for general relativity,''
  New J.\ Phys.\  {\bf 10} (2008) 083006
  [arXiv:0802.0864 [gr-qc]].



\bibitem{newangle}
 B.~Bahr and B.~Dittrich,
  ``Regge calculus from a new angle,''
  New J.\ Phys.\  {\bf 12} (2010) 033010
  [arXiv:0907.4325 [gr-qc]].
%



\bibitem{dittrichryan}
  B.~Dittrich and J.~P.~Ryan,
  ``Phase space descriptions for simplicial 4d geometries,''
  Class.\ Quant.\ Grav.\  {\bf 28} (2011) 065006
  [arXiv:0807.2806 [gr-qc]].
%


\bibitem{cylindrical}
B.~Dittrich,
  ``From the discrete to the continuous: Towards a cylindrically consistent dynamics,''
  New J.\ Phys.\  {\bf 14} (2012) 123004
  [arXiv:1205.6127 [gr-qc]].


\bibitem{zapata4d}
H.~Waelbroeck and J.~A.~Zapata,
  ``A Hamiltonian formulation of topological gravity,''
  Class.\ Quant.\ Grav.\  {\bf 11} (1994) 989
  [arXiv:gr-qc/9311035].
J.~A.~Zapata,
  ``Topological Lattice Gravity Using Self-Dual Variables,''
  Class.\ Quant.\ Grav.\  {\bf 13} (1996) 2617
  [arXiv:gr-qc/9603030].




\bibitem{dittrichryan23}
 B.~Dittrich and J.~P.~Ryan,
  ``Simplicity in simplicial phase space,''
  Phys.\ Rev.\ D {\bf 82} (2010) 064026
  [arXiv:1006.4295 [gr-qc]].
 B.~Dittrich and J.~P.~Ryan,
  ``On the role of the Barbero-Immirzi parameter in discrete quantum gravity,''
  arXiv:1209.4892 [gr-qc].

\bibitem{twisted}
L.~Freidel and S.~Speziale,
  ``Twisted geometries: A geometric parametrisation of SU(2) phase space,''
  Phys.\ Rev.\ D {\bf 82} (2010) 084040
  [arXiv:1001.2748 [gr-qc]].

\bibitem{williamsarea}
 J.~W.~Barrett, M.~Rocek and R.~M.~Williams,
  ``A Note on area variables in Regge calculus,''
  Class.\ Quant.\ Grav.\  {\bf 16} (1999) 1373
  [gr-qc/9710056].

  \bibitem{aristide3D}
  A.~Baratin and L.~Freidel
  ``Hidden Quantum Gravity in 3-D Feynman diagrams,''
  Class.\ Quant.\ Grav.\  {\bf 24} (2007) 1993
  [gr-qc/0604016].




\bibitem{lincoarse}
  B.~Bahr, B.~Dittrich and S.~He,
  ``Coarse graining free theories with gauge symmetries: the linearized case,''
  New J.\ Phys.\  {\bf 13} (2011) 045009
  [arXiv:1011.3667 [gr-qc]].






\bibitem{carloorient}
 M.~Christodoulou, M.~Langvik, A.~Riello, C.~Roken and C.~Rovelli,
  ``Divergences and Orientation in Spinfoams,''
  Class. Quantum Grav. 30 055009 2013
  [arXiv:1207.5156 [gr-qc]].


\bibitem{bojowald}
 M.~Bojowald,
  ``Loop quantum cosmology,''
  Living Rev.\ Rel.\  {\bf 8} (2005) 11
  [gr-qc/0601085].
 A.~Ashtekar and P.~Singh,
  Class.\ Quant.\ Grav.\  {\bf 28} (2011) 213001
  [arXiv:1108.0893 [gr-qc]].
  
    \bibitem{steinhausM} 
  B.~Dittrich and S.~Steinhaus,
  ``Path integral measure and triangulation independence in discrete gravity,''
  Phys.\ Rev.\ D {\bf 85}, 044032 (2012)
  [arXiv:1110.6866 [gr-qc]].



\bibitem{recursion}
 V.~Bonzom, E.~R.~Livine and S.~Speziale,
  ``Recurrence relations for spin foam vertices,''
  Class.\ Quant.\ Grav.\  {\bf 27} (2010) 125002
  [arXiv:0911.2204 [gr-qc]].



\bibitem{sorkin}
 R.~Sorkin,
  ``The Electromagnetic field on a simplicial net,''
  J.\ Math.\ Phys.\  {\bf 16} (1975) 2432
   [Erratum-ibid.\  {\bf 19} (1978) 1800].


\bibitem{dsl}
B.~Dittrich, L.~Freidel and S.~Speziale,
  ``Linearized dynamics from the 4-simplex Regge action,''
  Phys.\ Rev.\  D {\bf 76} (2007) 104020
  [arXiv:0707.4513 [gr-qc]].








\end{thebibliography}

}
\end{document}